# Control of domain states in rhombohedral PZT films via misfit strains and surface charges

I.S. Vorotiahin, A.N. Morozovska, E.A. Eliseev, Y.A. Genenko

Technical University of Darmstadt, Institute of Materials Science, Darmstadt, Germany

Institute of Physics NAS of Ukraine, Kyiv, Ukraine

Institute for Problems of Materials Science NAS of Ukraine, Kyiv, Ukraine

## Abstract
Using the Landau-Ginzburg-Devonshire theory, an influence of the misfit strain and surface screening charges, as well as the role of the flexoelectric effect, have been studied by numerical modelling in the case of a rhombohedral lead zirconate-titanate ferroelectric/ferroelastic thin film with an anisotropic misfit produced by a substrate. It was established that the magnitude and sign of the misfit strain influence the domain structure and predominant directions of the polarization vector, providing misfit-dependent phases with different favourable polarization components. Whilst strong enough compressive misfit strains favour a phase with an orthorhombic-like polarization directions, strong tensile misfits only yield in-plane polarization components. The strength of surface screening is seen to condition the existence of closure domain structures and, by increasing, supports the single-domain state depending on the value of the misfit strain. The flexoelectric effect exhibits a weak influence on the phase diagram of multi-domain states when compared with the phase diagram of single-domain states. Its effect, however, becomes significant in the case of skyrmion topological states, which spontaneously form near the film surface when compressive misfit strains are applied. Cooperative influence of the misfit strain, surface screening charges and temperature can set a thin rhombohedral ferroelectric film into a number of different polar and structural states, whereby the role of the flexoelectric effect is pronounced for topologically nontrivial structures.

## 1. Introduction

Thin films of lead zirconate-titanate (PZT) belong to the most widely used ferroelectric materials, having excellent piezoelectric properties and phase diagrams that allow using their whole potential in electronic devices [1, 2, 3, 4, 5, 6, 7]. Their properties are mostly well-known and thoroughly analysed [5, 8, 9, 10, 11, 12, 13]. In comparison with other phase symmetries, rhombohedral thin films of PZT are known to exhibit extra high piezoelectric properties [14]. However, understanding of their important features is still lacking, particularly, of the role of the flexoelectric effect, which is especially pronounced in thin films.

To control the ferroelectric polar states of thin films the influence of the substrate-induced misfit strain, flexoelectricity, and surface screening charges on the phase formation should be considered in conjunction with each other. This can be done by analysing strain-dependent phase diagrams obtained from numerical and analytical calculations within the framework of the Landau-Ginzburg-Devonshire (LGD) theory [15, 16]. The influence of the misfit strain caused by the substrate on the properties of the film is a cornerstone of strain engineering [17]. It can increase or suppress polarization [18], favour one structural phase over another [19], and stabilize certain domain structures [20].

Flexoelectric effect, though being weak in the regions with small gradient of elastic or/and polar properties, gains greater impact, when dealing with thin films, and can be critically important for the physical properties [21, 22]. An inherent property of virtually all solid materials, this electromechanical effect binds strain gradients and polarization in case of the direct flexoelectric effect and polarization gradient with elastic strain or stress in the conversed effect case. The most natural regions for the flexoelectricity to manifest itself in ferroelectric and ferroelastic structures are domain walls [23, 24], structural defects [25], and strained (e.g., due to the substrate mismatch) surfaces creating strain gradients across the films [26, 27, 28]. Ferroelectric thin films are suitable for detecting flexoelectricity for several reasons. They have high dielectric permittivity and molecular structures with mobile ions, displacement of which is responsible for spontaneous polarization occurrence, which also enhances the flexoelectric effect. The small-size scales of thin films also facilitate the detection of flexoelectricity, as it influences macroscopic properties stronger. Flexoelectricity affects such properties as dielectric permittivity [29, 30], critical thickness of ferroelectricity [31] and domain-wall conductivity [32, 33]. It shifts ferroelectric hysteresis loops by voltage, creating a so-called built-in potential [34, 35, 36, 37] (or even switches polarization actively [38, 39, 40]), generates morphotropic phase boundary (MPB)-like rotations of the spontaneous polarization [41], and invokes other effects like polarization induced bending [42], modification of phonon dispersion curves [43], *etc*. When thin films are epitaxially grown on rigid substrates with a mismatch, the misfit strains may be relieved via formation of local misfit dislocations, which, being inhomogeneous, also promote flexoelectricity [44]. This all means that accounting the flexoelectric effect when studying thin ferroelectric films becomes crucial.

Experimental studies and theoretical modelling of the misfit strain effects have their goal in establishing how the chemical composition and structural properties of specific substrates influence properties of the ferroelectric films being grown on them [45, 46]. Choosing specific combinations of a substrate and a film defines the value and direction of the misfit strain which in turn forms a specific structural phase [47], domain patterns [48, 49, 50], the order of phase transition, or can stabilize polymorphic ferroelectric phases with a coexistence of several structural phases in a way similar to the MPB [51, 52, 53, 54, 55]. In some cases, selected misfit strains help stabilize ferroelectricity in the otherwise paraelectric materials [56, 57, 58, 59, 60]. This is where both well-known [61] and novel [19] materials have a great potential. Generally, domain structures of the epitaxial thin films can be conveniently tailored by misfit engineering and could even be used to develop new types of devices with a domain wall as a functional element [45, 62, 63, 64, 65, 66, 67, 68].

Thin ferroelectric films of various materials, from PZT to $BiFeO_3$ (BFO) were previously modelled in different structural phases with the Landau-Ginzburg approach [69]. A phase-field study in 2005 [18] has shown that domain morphologies in tetragonal, rhombohedral and orthorhombic PZT films depend on the sign (tensile or compressive) of misfit. Thereby rhombohedral domain period changes due to a small misfit of ±0.5%, while tetragonal structures reconfigure with different polarization components depending on the strength of the tensile misfit. A similar research was conducted later for rhombohedral BFO films [70], where the sign and value of strain applied by the substrate, as well as the orientation of the film on the substrate, determine shapes, sizes and the very existence of domains in a thin film. A step further was a simulation of lead titanate (PTO) nanoparticles of different shapes and the impact of their geometry on phase transition parameters (e.g., Curie temperature) with occurrence of vortex domains [71]. Phase transitions conditioned by epitaxial strain were found and studied by density-functional theory [72]. Employing strains in the range of ±1.5%, phase behaviours of barium titanate (BTO) and PTO have been compared and the path of changes between several structural phases with different dominating polarization components was drawn. For BTO, the compressive strain inhibits and even suppresses the transition to the nonpolar phase [73]. The aforementioned appearance of the ferroelectricity in $SrTiO_3$ under misfit strains was also predicted [74] and investigated phenomenologically [75]. The mentioned theoretical studies reveal a great deal about the effect of the mismatch strains. They, however, did not account for the flexoelectric effect, the influence of which becomes significant at small scales. The general framework for the modelling studies consisted of efforts to develop multiscale simulation strategies that connect different theoretical and modelling approaches across a wide range of scales in order to realize a concept of "ferroelectric oxides by design" [45]. Most of the studies, be it theoretical or experimental, focus themselves on the (001)-oriented ferroelectric films due to ease of growth and commercial accessibility of the (001)-cuts [45].

Surface screening charges present another factor responsible for a variety of effects in ferroelectrics. Induced by surface defect states or an electrode, they affect the depolarizing field that influences a domain structure in the film. When the screening is close to the ideal one, it supports a single-domain state or at least opposes any sort of closure domain structures [76, 77, 78, 79, 80, 81]. Even the non-ideal screening affects polar properties of a ferroelectric film, typically having an impact on the nucleation process, domain shapes and sizes, polarization rotation, broadening of domain walls *etc.* [82, 83, 84, 85, 86, 87, 88, 89, 90, 91, 92, 93, 94, 95]. Thus, it is important to keep the surface screening in view or take it into account when studying properties of ferroelectric thin films.

The subject of this study is a single-crystalline film of $Pb(Zr_{0.6}Ti_{0.4})O_3$ (PZT 60/40), which serves as a model material to investigate properties of rhombohedral ferroelectrics on the nanoscale using the phenomenological concept and simulations. The study is performed by means of the time-dependent LGD theory with a system of equations incorporating important electrostatic and mechanical interactions, including electrostriction and the flexoelectric effect. The model and its constitutive equations are presented in Section 2. Section 3 displays the modelling results from two main perspectives: analytical consideration of the single-domain states and finite-element modelling (FEM) simulations of the multi-domain states of the film. The chart of single-domain states in terms of temperature and misfit strain is shown in Section 3.1. The multi-domain spatial distributions of electrostatic and mechanical quantities and their characteristics in dependence of the misfit strain are presented in Section 3.2. The topological polarization formations at the top surface of the film are considered in Section 3.3. The influence of the surface screening is investigated in Section 3.4. It is followed by a general discussion in Section 4 and Conclusions.

## 2. Theory and Methods
### 2.1. Problem Layout

Considered is a thin ferroelectric film of thickness $h$, attached to a bottom metallic electrode and exposed to the ambience from the top side. Up above, in the ambience, on a distance $h_v$ from the top surface, there is a top metallic electrode. Both electrodes are kept at fixed potentials. The thin film consists of PZT 60/40, which has a rhombohedral crystalline structure at room temperature. In the absence of any misfit strains between the substrate and the film, a domain structure spontaneously forms with time. Material parameters for the PZT 60/40 used here can be found in **Appendix A**.

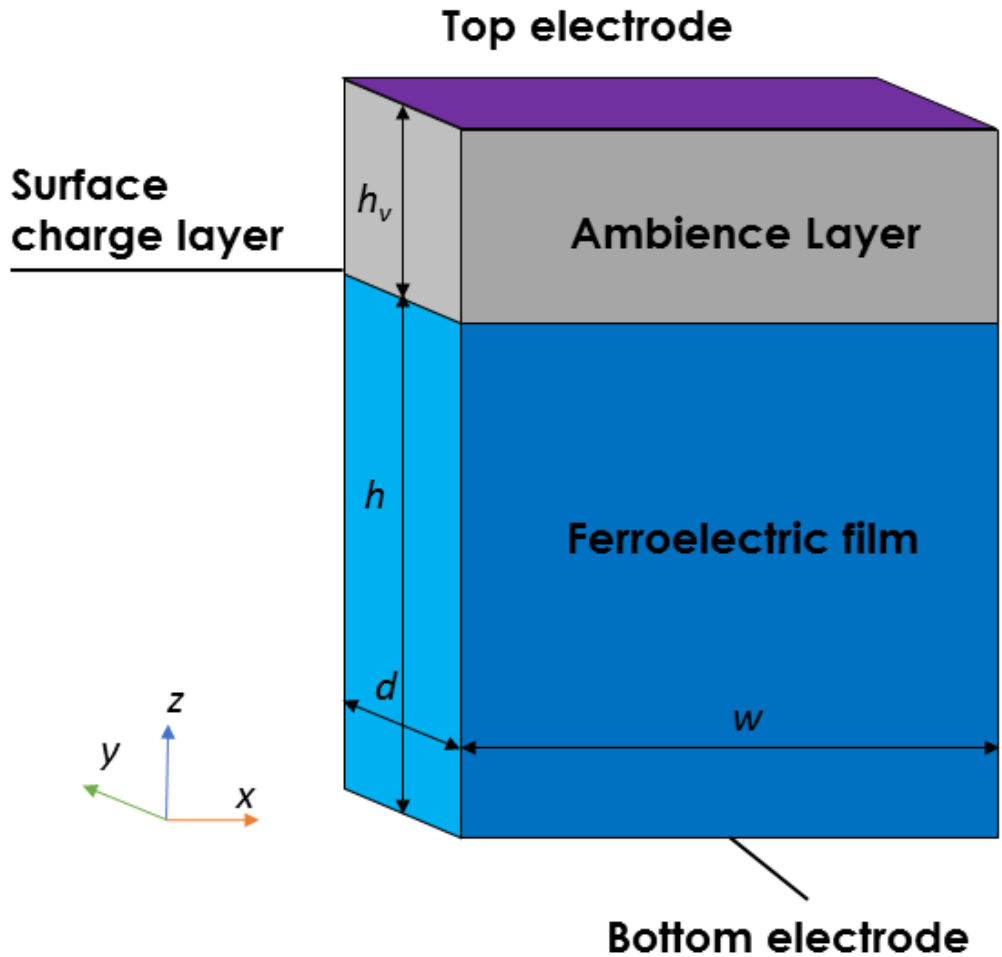

**FIGURE 1**. Problem layout. A ferroelectric film of thickness $h$, width $w$ and depth $d$ on the bottom electrode with an ambience layer of thickness $h_v$ on the top. A layer of screening charge on the top surface of the film separates it from the ambience. Here and afterwards, a correspondence shall be assumed between $x = x_1$, $y = x_2$, and $z = x_3$. Periodic boundary conditions along $x$ and $y$ extend the film infinitely in the $(x, y)$ plane.

## 2.2. Gibbs Free Energy Functional

The Gibbs thermodynamic potential of the whole system can be described as a sum of a volume free energy, a surface free energy, and an ambience free energy:

$$G = G_V + G_S + G_A. \tag{1}$$

In relation to polarization **P**, electrostatic potential $\varphi$ and elastic stress $\sigma_{ij}$, the volume, surface and ambient parts of the Gibbs free energy respectively read as follows:

$$G_V = \int_V d^3r \left( \begin{array}{c} \frac{\alpha_{ik}}{2} P_i P_k + \frac{\beta_{ijkl}}{4} P_i P_j P_k P_l + \frac{\gamma_{ijklmn}}{6} P_i P_j P_k P_l P_m P_n + \frac{g_{ijkl}}{2} \left( \frac{\partial P_i}{\partial x_j} \frac{\partial P_k}{\partial x_l} \right) \\ -P_i E_i - Q_{ijkl} \sigma_{ij} P_k P_l - \frac{s_{ijkl}}{2} \sigma_{ij} \sigma_{kl} - F_{ijkl} \sigma_{ij} \frac{\partial P_l}{\partial x_k} \end{array} \right), \tag{2a}$$

$$G_S = \int_S \left( \frac{\alpha_{ij}^S}{2} P_i P_j - \frac{\varepsilon_0}{2\lambda} \varphi^2 \right) d^2r, \tag{2b}$$

$$G_A = -\int_A \frac{\varepsilon_0 \varepsilon_v}{2} E_i E_j d^3r. \tag{2c}$$

As mentioned in Table I, the coefficients of the LGD expansion by the powers of polarization are $\alpha_{ik}$, $\beta_{ijkl}$ and $\gamma_{ijklmn}$; $g_{ijkl}$ is the gradient coefficient tensor. $E_i$ is the electric field component; $\sigma_{ij}$ is the elastic stress tensor; $Q_{ijkl}$ is the electrostriction tensor; $F_{ijkl}$ is the flexoelectric effect tensor; $s_{ijkl}$ is the elastic compliance. $\alpha_{ij}^S$ is a surface stiffness tensor, here and after presumed to be zero; $\lambda$ is the surface screening length parameter for the Bardeen model of surface charges. $\varepsilon_0$ is the dielectric constant for the vacuum, and $\varepsilon_v$ is a relative dielectric constant of the ambience. Relations of the tensor coefficients with the rationalized (Voigt) matrix notations often used in literature and also below throughout this work are listed in **Appendix A**.

## 2.3. System of Equations

Equations solved in this study include the time-dependent Landau-Khalatnikov-type LGD equations for three polarization components $P_i$, the Poisson equation for the electrostatic potential $\varphi$, and the equilibrium condition for elastic stress components $\sigma_{ij}$.

### Coupled time-dependent LGD Equations

The coupled time-dependent LGD equations can be obtained by the variation of the Gibbs free energy with respect to the order parameter of polarization,

$$\Gamma \frac{\partial P_i}{\partial t} = -\frac{\delta G_V}{\delta P_i}. \qquad (i = 1, 2, 3) \tag{3}$$

The equations read explicitly as follows:

$$\Gamma \frac{\partial P_k}{\partial t} + \alpha_{ik} P_i + \beta_{ijkl} P_i P_j P_l + \gamma_{ijklmn} P_i P_j P_l P_m P_n - g_{ijkl}\left(\frac{\partial^2 P_i}{\partial x_j \partial x_l}\right) - 2Q_{ijkl}\sigma_{ij}P_l + F_{ijkl}\frac{\partial \sigma_{ij}}{\partial x_l} = -\frac{\partial \varphi}{\partial x_k}. \quad (4)$$

Boundary conditions for top and bottom surfaces include flexoelectric coupling

$$\left(g_{ijkl}\frac{\partial P_i}{\partial x_j} - F_{ijkl}\sigma_{ij}\right)\bigg|_{x_3=0,h} = 0 . \quad (5)$$

### Poisson Equation

The electrostatic properties of the material can be described using the Poisson equation for the electric potential φ related to the electric field, $E_i = -\nabla_i \varphi$. It contains terms correspondent to the contributions from the crystal lattice itself and ferroelectric polarization. The spatial distribution of φ can be devised both for the film bulk and for the ambience, where the ferroelectric contribution is absent. Thus, for the film the Poisson equation reads as

$$\varepsilon_0 \varepsilon_b \frac{\partial^2 \varphi}{\partial x_i \partial x_i} = \frac{\partial P_j}{\partial x_j}, \quad (6a)$$

while in the ambience it turns into the Laplace equation,

$$\varepsilon_0 \varepsilon_v \frac{\partial^2 \varphi}{\partial x_i \partial x_i} = 0 \quad (6b)$$

with boundary conditions setting up the properties of the bottom electrode (ground), the interface between the film and the ambience, and the top electrode:

$$\varphi|_{x_3=0} = 0, \ \left(D_n^{ext} - D_n^{int} + \varepsilon_0 \frac{\varphi}{\lambda}\right)\bigg|_{x_3=h} = 0, \ \varphi|_{x_3=h-0} = \varphi|_{x_3=h+0}, \ \varphi|_{x_3=h+d} = 0, \quad (7)$$

where the electric displacement is defined by $\mathbf{D}=\varepsilon_0\varepsilon_b\mathbf{E}+\mathbf{P}$, and λ is an effective screening length [96, 97].

### Generalized Hooke's Law

Mechanical and electromechanical properties of the film are taken into account via the generalized Hooke law that describes relationship between strain and stress within the ferroelectric film with the effects of interest being included, such as electrostriction and flexoelectric effect ("flexoeffect" for brevity). It can be derived by variation of the Gibbs free energy (2a) with respect to stress variable $\sigma_{ij}$,

$$u_{ij} = s_{ijkl}\sigma_{kl} + F_{ijkl}\frac{\partial P_k}{\partial x_l} + Q_{ijkl}P_k P_l . \quad (8)$$

In the system of equations, elastic stresses are governed by an equilibrium condition, namely:

$$\frac{\partial \sigma_{ij}}{\partial x_i} = 0 \quad (9)$$

It is solved with the following (mixed) boundary conditions on elastic strain $u_{ij}$ and stress $\sigma_{ij}$:

$$u_{11}|_{x_3=0} = u_m, \; u_{22}|_{x_3=0} = 0, \; u_{33}|_{x_3=0} = 0; \; \sigma_{11}|_{x_3=h} = 0, \; \sigma_{22}|_{x_3=h} = 0, \; \sigma_{33}|_{x_3=h} = 0, \qquad (10)$$

where $u_m$ is the misfit strain.

The system of equations described above has been solved with the finite element method using COMSOL@Multiphysics software package. To observe the formation of the domain structure in a part of the ferroelectric film we have enclosed the region inside the computational box with dimensions of $w \times d \times (h + h_v)$ along, respectively, $x_1$, $x_2$ and $x_3$ (or, respectively, $x$, $y$ and $z$) axes. Periodic boundary conditions were imposed to the boundaries along $x_1$ and $x_2$ axes. The dimensions $w, d, h$, of the computational box are: $w = 24$ nm, $d = (8 - 24)$ nm, $h = 24$ nm, $h_v = 12$ nm.

## 3. Results

### 3.1. Phase diagram of single-domain states

To support the analysis of spatial polarization distributions and domain structures to be derived from the numerical solution of Eqs. (4,6,9) we first evaluate the phase diagram of single-domain states of the film in dependence of the misfit strain and temperature by analytical solution of the LGD equation system (see **Appendix B.a.**).

Figure 2 presents a phase diagram for a single domain state of a thin PZT 60/40 film with three polar phases and one nonpolar phase dependent on temperature and *uniaxial* misfit. The polar phases have the electric polarization oriented in the in-plane ($P_1$, $P_2$), out-of-plane ($P_3$), and mixed directions. The high-temperature phase is nonpolar at small and compressive misfit strains. The out-of-plane polarization ($P_3$) dominates at the compressive misfit strains and gradually fades away with increasing tensile strains. The in-plane component $P_1$ is, on the contrary, dominant at tensile misfit strains and fades away when they become compressive. In case of isotropic *biaxial* misfit strain (see **Appendix B.b.**), the other in-plane polarization component $P_2$ reproduces the behaviour of $P_1$, and creates an out-of-plane-dominant phase at high compressive misfit strains, an in-plane-dominant phase at high tensile misfit strains, and a mixed ($P_1$, $P_2$, $P_3$) rhombohedral phase between them at small misfit strains.

In the case of anisotropic *uniaxial* misfit strain (Fig. 2), the component $P_2$ is persistent throughout almost entire misfit range at lower temperatures. It creates a rich variety of polar phases. The persistent $P_2$ phase intersects the paraelectric phase, resulting in a small range, where the $P_2$ component is dominant. At higher compressive misfit strains, $P_2$-phase is partially combined with a $P_3$-dominant phase, creating a separate phase, where polarization has both $P_3$ and $P_2$ components. At higher tensile misfit strains, it creates a band where both $P_1$ and $P_2$ can be realized, before the film goes into the $P_1$-dominant state. The bottom section of a diagram at lower misfits and lower temperatures represents a phase, where all three components of the polarization vector are nonzero. Dotted lines show how would the phase boundaries change, if it were not a thin film, but a thick ferroelectric bulk (for details see **Appendix B.b.**).

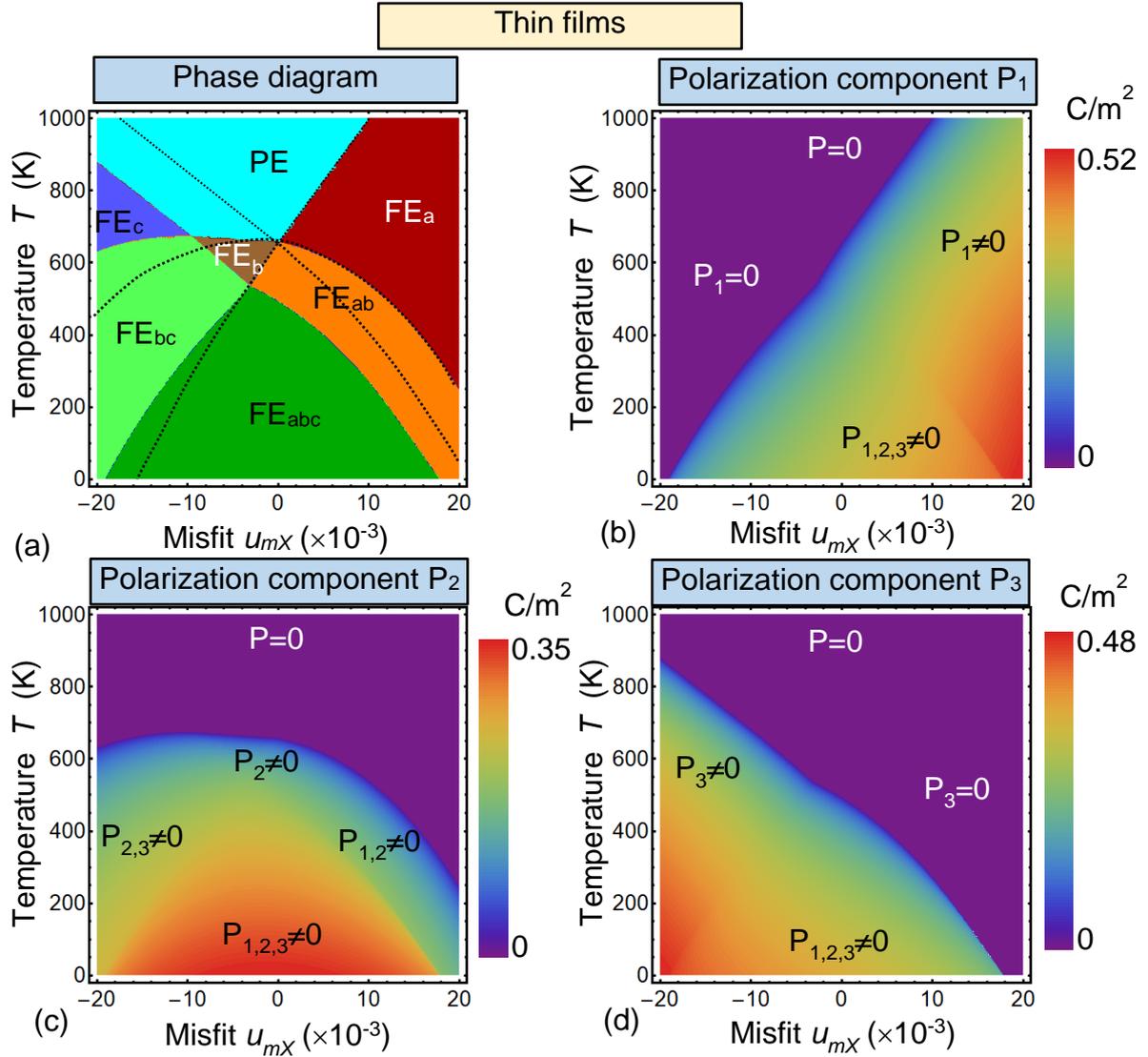

**FIGURE 2**. **(a)** Ferroelectric PZT 60/40 phase diagram in coordinates temperature – misfit strain for thin films. Anisotropic uniaxial misfit strain between the film and substrate is considered. Polarization components $P_1$ **(b)**, $P_2$ **(c)** and $P_3$ **(d)** dependence on the misfit strain and temperature. Color bars show the range of corresponding values for the contour maps. Paraelectric phase is denoted as "PE", while "FE$_a$", "FE$_b$", "FE$_{ab}$", "FE$_{abc}$", "FE$_{bc}$", and "FE$_c$" show the regions for ferroelectric phases with different orientation of polarization, namely, "a" means $P_1$, "b" means $P_2$ and "c" means $P_3$. Film thickness is $h$=24 nm and effective dielectric gap width is $d_{eff}$=0.12 nm (see **Appendix B.a.**). Bulk parameters are listed in **Table BI** in **Appendix B**. Here we suppose that $u_1^{(m)} \equiv u_{mX}$, while $u_2^{(m)} = 0$ and $u_6^{(m)} \equiv 0$.

### 3.2. Domain structures

A series of spatial distributions has been obtained as a result of 3D FEM simulations at room temperature using COMSOL@Multiphysics software. Among those, of interest were the polarization distributions, which illustrate the domain structure that should naturally occur in PZT 60/40 films, as well as mechanical displacement distributions that show a deformation pattern of the film under given conditions. The map of polarization vectors is presented exemplarily in Fig. 3 for the case of zero misfit $u_m = 0$ where arrow directions show the direction of polarization at a given point, and an arrow length represents a relative amplitude of polarization. The color scale from blue to red indicates variation of the *z*-component of polarization between its lowest and highest values to easier identify the domain patterns. This allows observation of a periodic domain arrangement along the longest box size in the *x*-direction as well as the closure domains in the free top surface of the film.

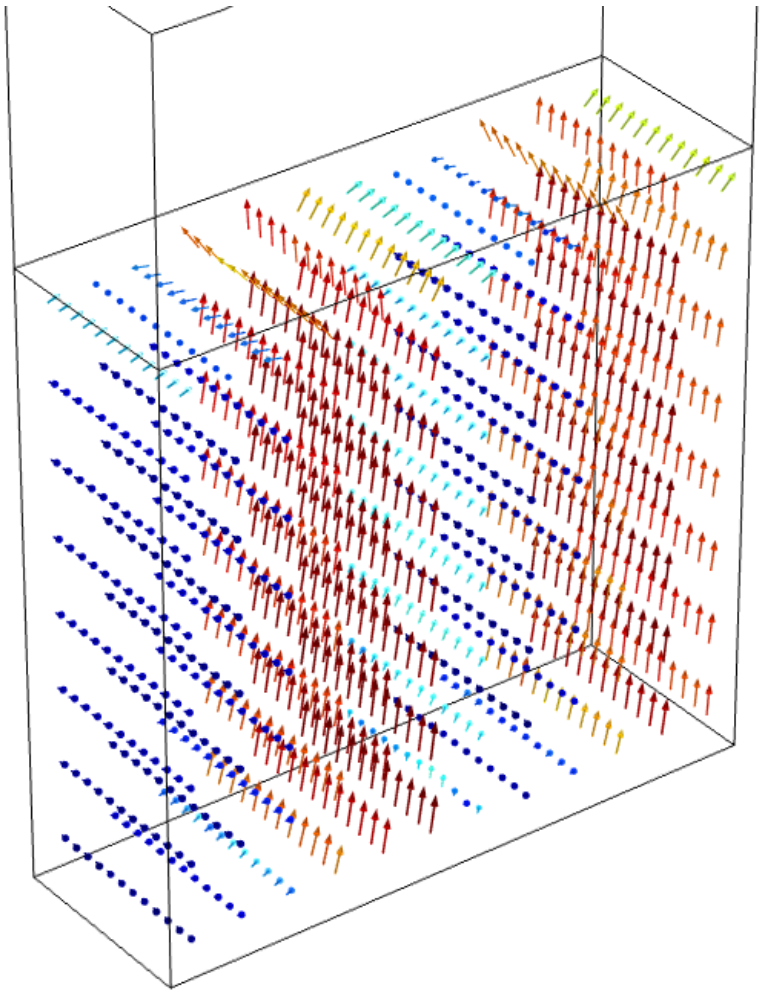

**FIGURE 3**. Arrow fields of the polarization vector inside a ferroelectric PZT film with flexoeffect and zero misfit strain.

To establish a role of the substrate, a series of computations has been performed, where the uniaxial misfit strain on the bottom surface of the film was varied between -2% and 2% with a step of 0.25%. In between the characteristic values of -0.5% and 0.5%, this step was reduced to 0.1%. The resulting structures vary significantly in directions of their polarization vectors, domain walls, as well as in deformations of films. To compare the domain structures directly the polarization maps are shown in Fig. 4 for different negative and positive misfit strains.

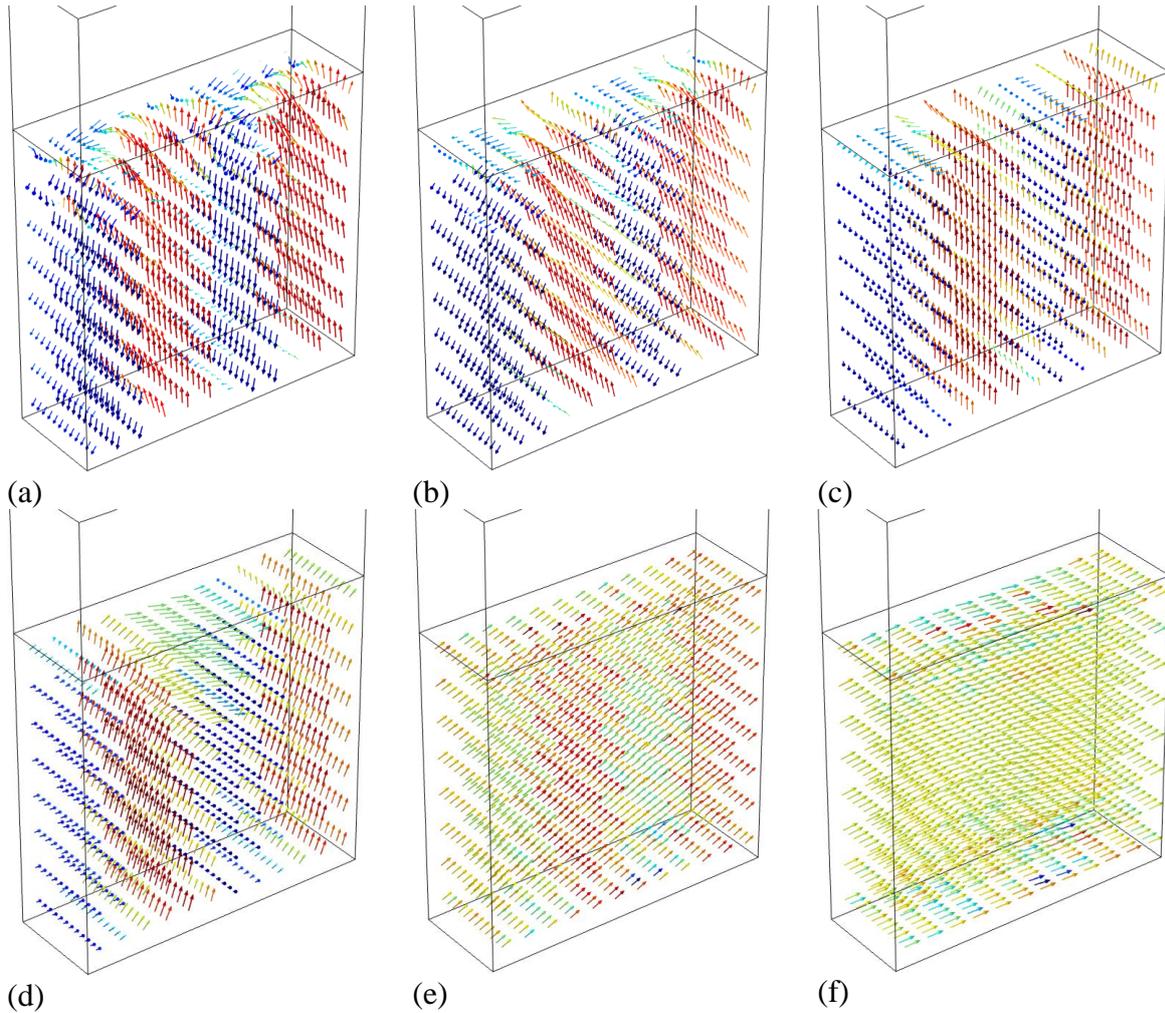

**FIGURE 4**. Arrow fields of the polarization vector inside a ferroelectric PZT film for misfit strains at the bottom electrode $u_m = -2\%$ (a), $-1\%$ (b), $-0.5\%$ (c), $0.5\%$ (d), $1\%$ (e) and $2\%$ (f).

All the polarization maps but the cases with the two highest tensile strains exhibit periodic domain structures whereby distinct effects of compressive and tensile misfit strains on the rhombohedral PZT film are revealed. The compressive misfit strain favours the $P_3$ polarization component and roughly vertical domain walls (Fig. 4(a-c)). At two highest values of the compressive strains (Fig. 4(a,b)), complicated three-dimensional closure polarization structures can be observed near the top surface of the film whose properties will be discussed below in Section.... The increasing tensile misfit strains (Fig. 4(d-f)) facilitate another behaviour, at which the film gradually loses its multi-domain structure. At the highest misfit strain (Fig. 4(f)), the film is predominantly polarized in the $P_1$ direction given by the uniaxial misfit.

Transformations of the domain structure can be better understood when observing the evolution of the absolute values of the polarization components averaged over the film volume as is shown in Fig. 5. The polarization vectors have all three components of roughly equal amplitudes, as expected in a rhombohedral phase, when the misfit strain is close to zero or weakly compressive. However, increasing compressive misfit strain in *x*-direction eventually reduces the $P_1$ component, so that below $u_m = -1\%$ polarization vectors are mostly tilted within the $(P_2, P_3)$-plane that is rather expected in an orthorhombic phase. The $P_3$ component exhibits an opposite behaviour. It is at maximum at the highest compressive misfit strain and decreases monotonically, when the latter changes from compressive to tensile strains, to virtually vanish at about $u_m = 1\%$. Above this value, a modulated in-plane polarization structure is formed which is discussed below. No significant effects of the presence or absence of the flexoelectric coupling were detected which only provide minor changes in domain shapes and precise form of film's deformation.

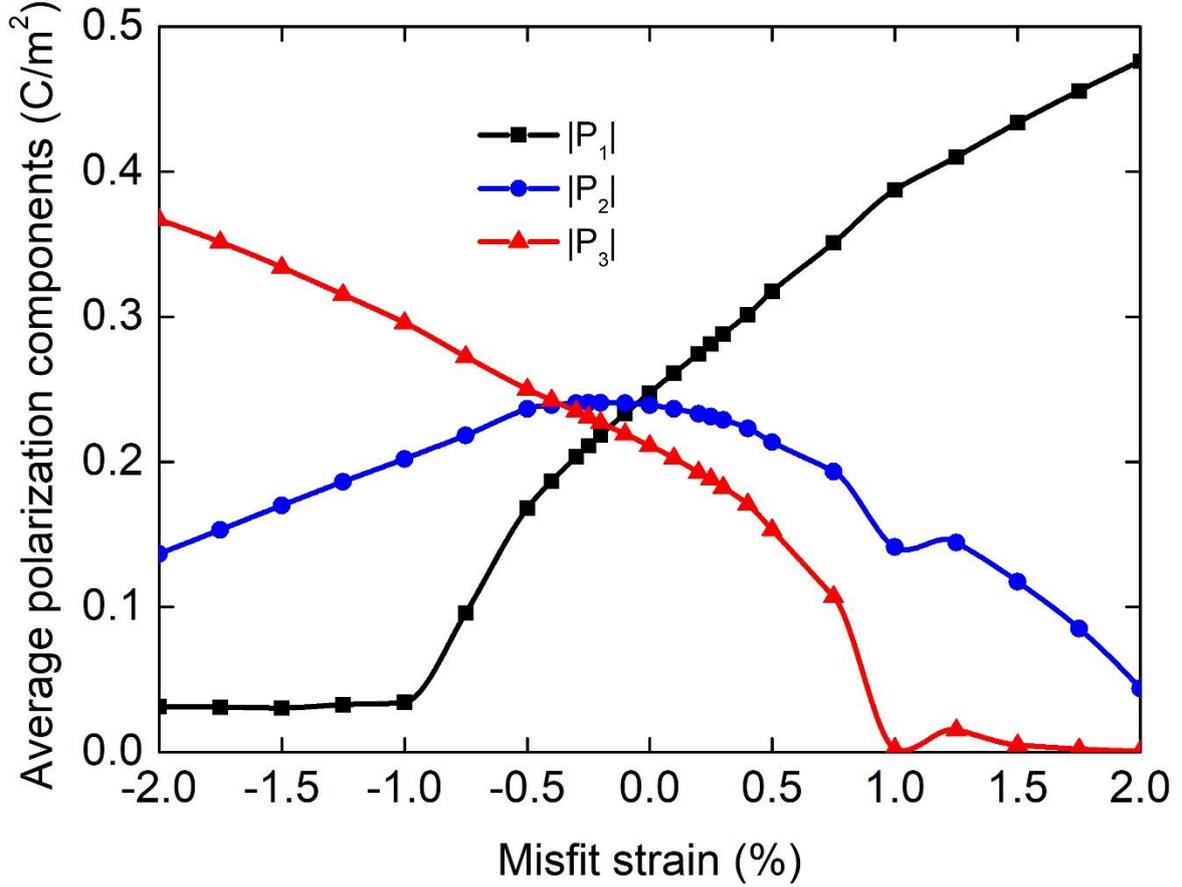

**FIGURE 5**. Misfit strain dependencies of the volume average of $P_1$, $P_2$, and $P_3$ components of polarization.

When increasing the tensile misfit over $u_m = 1\%$ a phase with in-plane polarization ($P_1 \neq 0, P_2 \neq 0$, see Fig. 6(a)) is stabilized forming thereby peculiar folds induced by modulation of the component $P_2$ in the film plane (see Fig. 6(b,c)). Here the calculations were performed in the absence of the flexoeffect and the presented deformation is upscaled for better visibility by factor 30. The amplitude of the $P_2$ modulation is smaller than the value of the approximately uniform component $P_1$ (see Fig. 6(a)). A threshold value for the appearance of the modulation is determined by the spontaneous deformation of the non-constrained sample. If the mean value $u_1^{(s)} = Q_{11}P_1^2 + Q_{12}P_1^2$ is smaller than the misfit strain $u_m$ (polarization value is also affected by the misfit strain), the film is extended but remains uniform. When $u_1^{(s)}$ exceeds $u_m$, a sort of phase transition occurs similar to the Euler instability in mechanical systems.

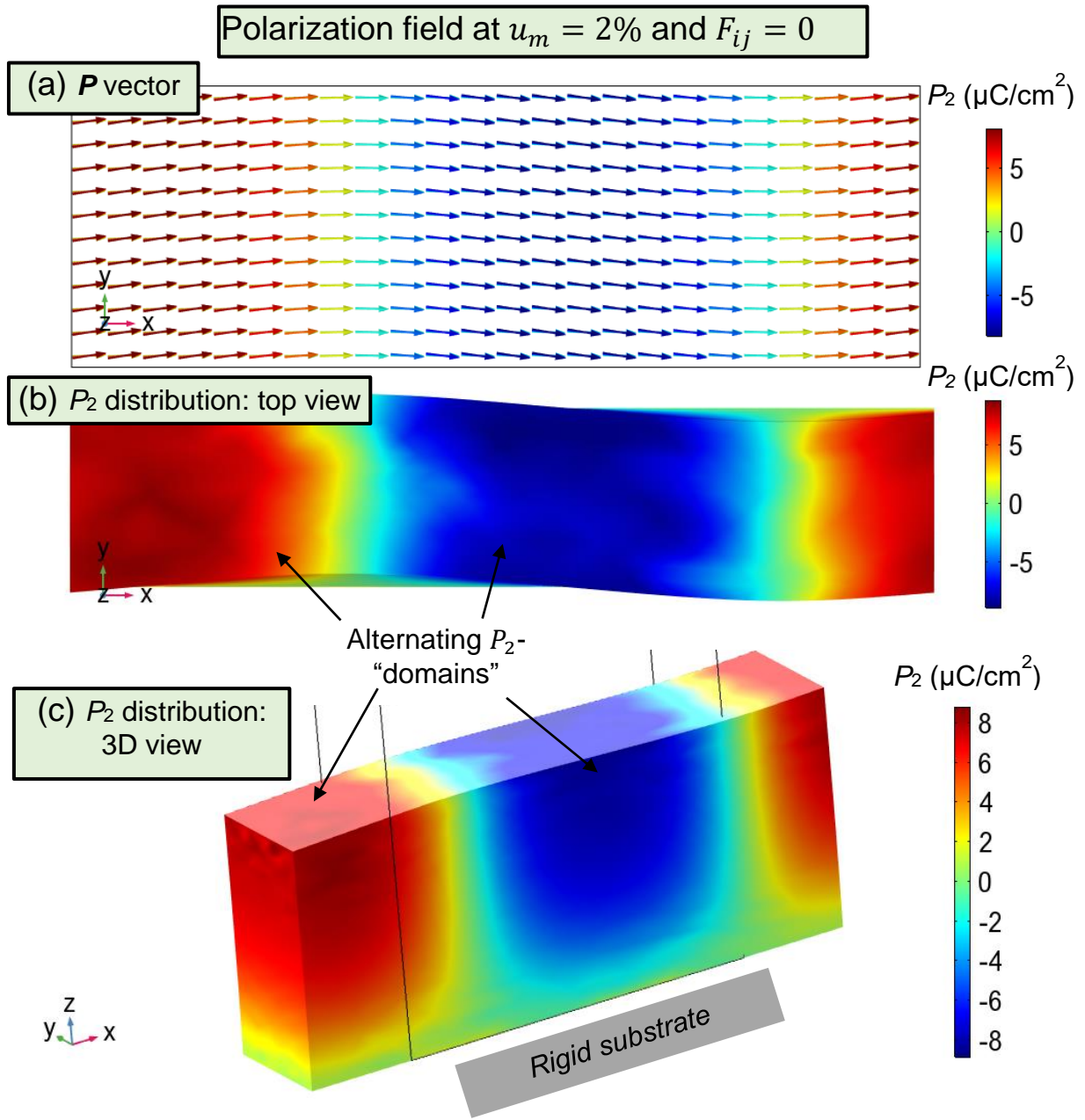

**FIGURE 6.** **(a)** Distribution of polarization vector **P** at the film top surface shown by arrows colored according to the value of component $P_2$. The distribution of the latter on the top free surface **(b)** and on the "surface" of computational cell **(c)**.

As is seen in Fig. 6(c), there is no polarization modulation near the film bottom ($P_2 = 0$), that is the film is firmly attached to the rigid substrate whose elastic moduli are assumed higher than those of the film. In this case, all displacements are determined by the misfit strain tensor components. Spreading of the "domain walls" near the bottom is caused by distinct boundary conditions at the interface film/bottom and the free top surface of the film (see e.g. [98])

The presented FEM results for multi-domain states can be compared with the single-domain phase diagram displayed in Section 3.1, if we draw a line going through the said diagram at 300 K. At no misfit and between the values from -1% to +1%, the mixed rhombohedral-like phase, where all three components have comparable values, is located on the diagram with the most thermally stable point at approximately -0.5% strain, that is in agreement with the multi-domain results of Fig. 5. When we proceed increasing the compressive misfit strain, the $P_1$ value will gradually weaken and vanish at the misfit strain of -1.1%, below which we only see $P_2$ and $P_3$ components remaining in the single-domain state. The multi-domain simulation reveals in this region a small but stable $P_1$ component. When increasing the tensile misfit strain, a similar situation can be observed with fading of the $P_3$ component, which vanishes at +1% misfit strain. Above this value the polarization vector has only $P_1$ and $P_2$ components in a good agreement between the two approaches up until $u_m = 1.9\%$, where $P_2$ also vanishes in the single-domain picture while it remains finite and modulated in the multi-domain simulations.

In spite of the mostly monotonic behaviour of polarization components across the studied misfit strain region, a phase transformation from the rhombohedral-like phase to the in-plane modulated phase, which occurs at $u_m = 1\%$, is clearly pronounced in the peak behaviour of the piezoelectric coefficients well visible in Fig. 7.

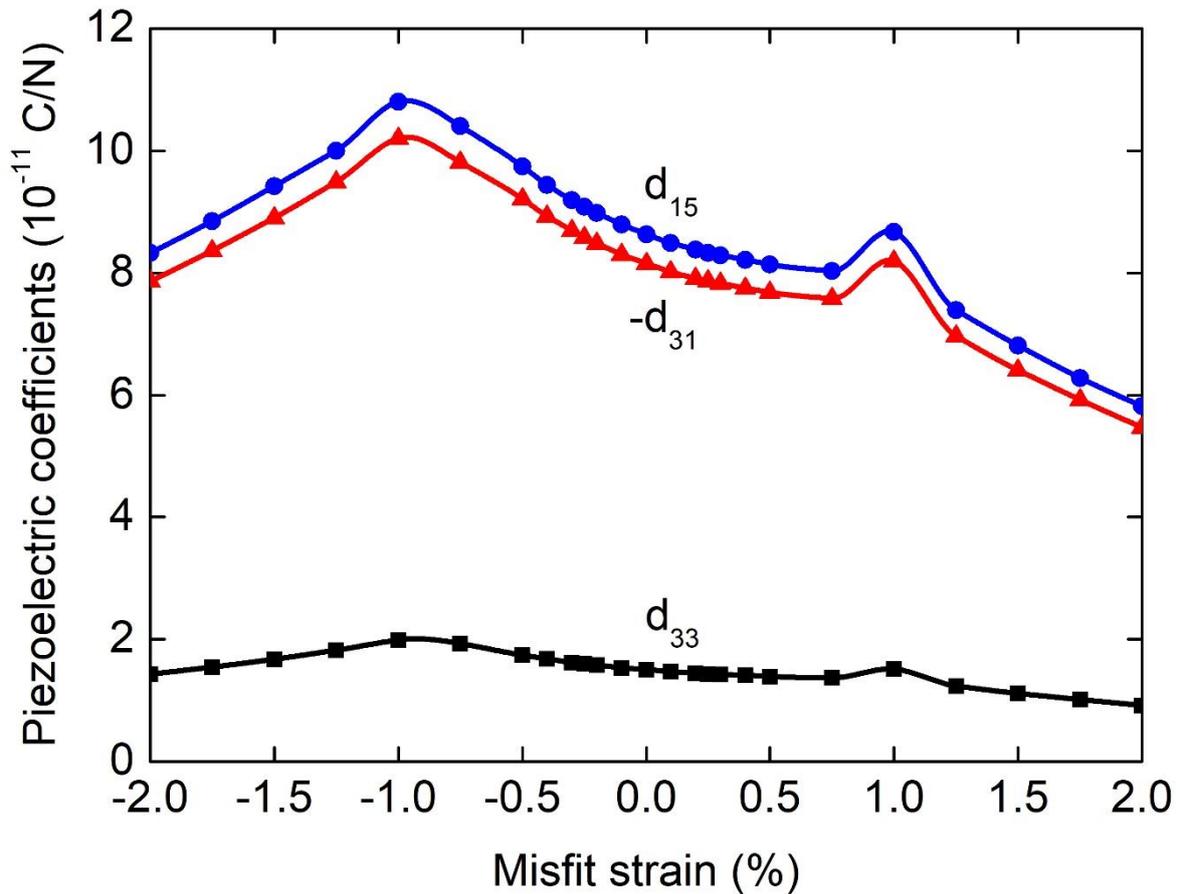

**FIGURE 7**. Dependence of the piezoelectric coefficients $d_{33}$, $d_{31}$ and $d_{15}$ of a 24-nm thick PZT 60/40 film on the substrate misfit strain.

### 3.3 Topologic polarization structures at the top film surface

Stable and reproducible three-dimensional polarization closure structures spontaneously formed near the film surface at high compressive misfit strains (see in Fig. 3(a,b)) remind of polarization vortices studied recently theoretically and experimentally in ferroelectric nanoparticles ([99, 71]) and thin films [86, 87]. They might also be related to chiral topological structures known in ferromagnetic materials [100, 101] and recently revealed theoretically [102, 103] and observed experimentally [104] in ferroelectrics. To check the possible chiral properties of structures in Fig. 3(a,b) we have calculated the topological index $n$ of these polarization configurations defined in terms of the unit polarization orientation vector $p$ as

$$n = \frac{1}{4\pi} \int_S \vec{p} \cdot \left[\frac{\partial \vec{p}}{\partial x} \times \frac{\partial \vec{p}}{\partial y}\right] dxdy. \tag{11}$$

The evaluation of the above integral was performed over the cross-section of the computational box in the $(x,y)$ plane located 1 nm below the top film surface. The index (11) can be of either sign and is supposed to take an integer (winding) number characterizing the vorticity of the object. The dependence of the index $n$ on the misfit strain is presented in Fig. 8 with and without account of the flexoelectric effect.

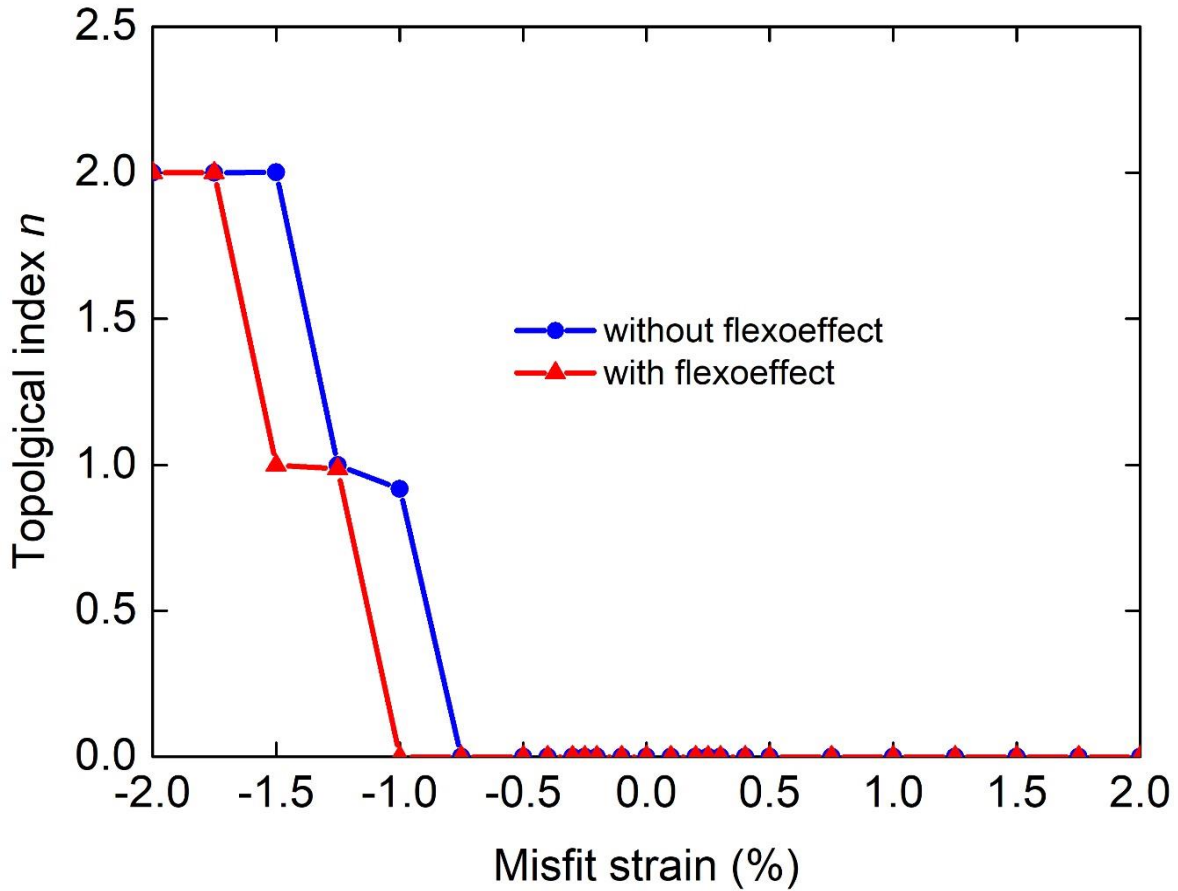

**FIGURE 8**. Dependence of the topological index $n$ characterizing polarization formations near the top film surface shown in Fig. 3 as a function the misfit strain.

Though it is believed that the energy contributions containing Lifshitz invariants, like those describing the flexoelectric coupling, facilitate modulated states with magnetization rotation and provide the stabilization of magnetic skyrmions [101], in the case of ferroelectrics, chiral topological formations surprisingly appear also without such contributions in the energy. Furthermore, in the absence of the flexoeffect, they exist in a wider region of compressive misfit strains (see Fig. 8). Thereby the index value $n = 2$ describes rather the presence of two chiral objects (see Fig. 4(a)) than the double vorticity of one object. The appearance of skyrmion-like formations in thin rhombohedral ferroelectric films under compressive misfit strains and the role of the flexoeffect in this phenomenon will be a subject of further detailed studies.

### 3.4 Surface Screening

Another area of interest when exploring the domain structures of rhombohedral ferroelectrics is the influence of the surface screening charge on their domain structures. Such physical effect in the current work is described by the linear Bardeen screening model and is represented by a term $\varepsilon_0 \varphi / \lambda$ in the boundary condition (7) for the Poisson equation. By varying parametric surface screening length $\lambda$ in this term, we can define strength of the surface screening charges, knowing that the screening is strong when $\lambda \ll 0.1$ nm and weak when $\lambda \gg 0.1$ nm [93]. Weak screening represents defects and impurities on the free surface of a film or charged particles sedimented on the surface. Strong screening, on the other hand, emulates an electrode put on the free surface. Such electrode is not fixed on the film, so that mechanically free conditions still apply at the top surface. Calculations here are performed with and without account of the flexoelectric effect but have proven to be identical for all practical purposes.

Studying domain structures that form in the presence of surface charges with the screening lengths between $10^{-3}$ nm and 10 nm, the changes in the domain structure were observed only in the range of $\lambda$ between $10^{-3}$ and $10^{-1}$ nm. Figure 9 shows a comparison of domain structures formed under misfit strains of -2, -0.5, 0, and 1.5 %, which represent all observed polar phases, at strong and weaker surface screening lengths. With gradually changing values of the surface screening length, the domain structure transits smoothly between the two states displayed for each column in the figure for exemplary strains. The main feature of the weaker screening is the presence of a closure domain structure underneath the top surface, meaning that the surface screening is too small to compensate the charges, produced by the domains at the top surface, and the structure needs to compensate it by changing itself. The structure formed under -2% compressive misfit strain exhibits the topological features discussed in the Section 3.3. These features, however, are not observed anymore at the strongest observed screening whereby only weak closure domains are formed. A general trend by variation from the compressive to the tensile misfit strains is the change from the vertical domain walls at $u_m = -2\%$ to deformed domain walls at $u_m = -0.5\%$, followed by an almost destroyed domain structure at $u_m = 0$ and finally by modulated in-plane polarization at $u_m = 1.5\%$. Thus, for the film subject to a tensile misfit (compare Figs. 9h and 9d), the changes caused by the surface screening strengthening are rather small. It makes an out-of-plane component more pronounced wherever a remnant domain with such orientation still exists, while the in-plane domains, occupying the vast majority of the bulk, hardly react to the surface screening.

The observed results show a rather predictable behaviour. Just like it was already established for tetragonal structures [86, 87, 95], the main function of surface screening charges is to compensate charges on the free surface generated by out-of-plane domains. Thus, at strong surface screening, there should be no closure domain structures, and it is not uncommon, but rather expected for the film to go for a single-domain structure as a stable or metastable state.

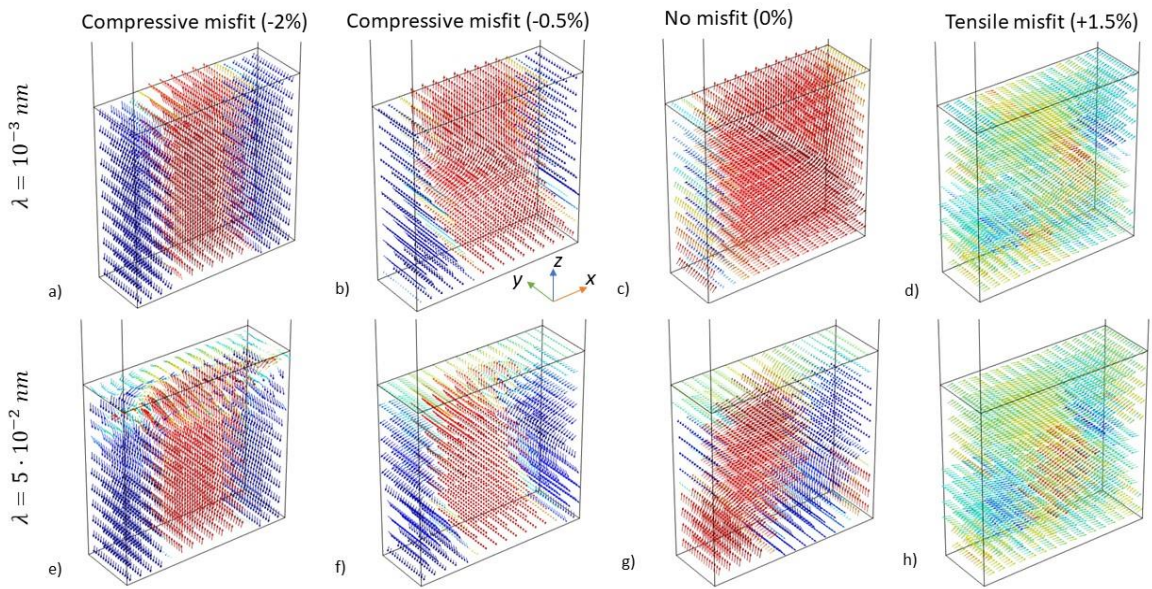

**Figure 9**. Comparison between domain structures in a thin PZT 60/40 rhombohedral ferroelectric film with varying from left to right misfit strains of -2% (a, e), -0.5% (b, f), 0% (c, g), and +1.5% (d, h) at stronger ($\lambda = 10^{-3}\ nm$) (upper row) and weaker ($\lambda = 5 \cdot 10^{-2}\ nm$) (lower row) surface screening.

# 4  Discussion

The performed modelling has shown how the rhombohedral thin ferroelectric films stand out in comparison to e.g., their tetragonal counterparts in terms of phase transformations and electromechanics. Dependent on the misfit strain caused by a substrate, domain structure in such films of PZT 60/40 can contain different types of domain walls and polarization directions in domains that can align by one, two or three components, forming regions with the rhombohedral-like, ac-orthorhombic-like and aa-tetragonal-like polarization patterns. This can be important for the electromechanical properties, because such characteristics like piezocoefficients are influenced by the number of dimensions in which polarization vectors rotate throughout the film volume (Fig. 7). Such characteristic is dependent on strains inside the film, and those can be caused or greatly modified by the mismatch with the substrate. At the detected points of ongoing structural changes (-1% and +1 % uniaxial misfit strains), we also see an increase in the piezocoefficient values, which further supports an idea that they indicate phase transitions. As the rhombohedral ferroelectric films are valued because of their relatively high piezocoefficients, the possibility to control them can be instrumental for electromechanical applications.

Comparison of the analytical Landau approach and FEM-LGD results demonstrates a principal agreement in a misfit-strain dependent phase range. The analytical results for single-domain states (Fig. 2) show the dominance of the in-plane polarization components for tensile misfit strains above 1%, whereas the FEM results (Fig. 5 and 6) without accounting for the flexoelectric effect further reveal the modulated in-plane ferroelectric phase.

Surface screening charges influence the domain structure in a predictable way. Similarly to the known cases of tetragonal materials [105, 86, 87, 95], strong screening favours the single-domain structure and suppresses perturbations. If a misfit strain is applied to the film, achieving the single-domain state is handicapped, but the destruction of closure domains is still predicted for the strong enough screening cases. This, however, counts for the domains with vertical polarization components. When and where the a-domains with only horizontal components are favoured by the strained system, the surface screening has no effect on their structure apart of eliminating small polarization perturbations. This tells us that, similarly to the tetragonal films, surface charges can be used to control domain structures of rhombohedral and, presumably, orthorhombic ferroelectric films.

An unexpected finding of the FEM-simulations is a formation of stable skyrmion-like topological polarization patterns at the free top surface of the film which are facilitated by high compressive strains and remain even in the absence of the flexoelectric effect as well as in the presence of a considerable screening.

# 5 Conclusions

The influence of uniaxial misfit strains, surface charges, and the flexoelectric effect on domain patterns and shape of a thin rhombohedral ferroelectric film was investigated by means of 3D finite element modelling using the Landau-Ginzburg-Devonshire approach. It was observed that spatial distributions of polarization and strain strongly change in dependence on the sign and value of the misfit strain and are conditioned by the presence of the flexoelectric effect. This means that a rhombohedral ferroelectric film can be manipulated into exhibiting different polar and elastic states via the substrate misfit, stabilizing different types of domain structures and deformation patterns. Components of polarization vectors, obtained for each studied misfit strain value, allow for formulating a diagram with structural phases, at 300K, beginning from the orthorhombic-like structure at higher compressive misfits going through the rhombohedral one around zero misfit into the tetragonal-like at higher tensile misfits. A particular position where the rhombohedral phase is the strongest, as evidenced by polarization vector components and deformation patterns, lies in the interval between -0.5 and 0% of the misfit strain. For the tensile strains, a transition to the spatially-modulated in-plane polar phase occurs at +1%. The flexoelectric effect, considered weak by itself, has a considerable impact on the structure in cases of the tensile misfit strains, particularly on the formation of skyrmion-like polarization patterns near the free top surface of the film. Surface screening charges are seen to be a tool for controlling polar states by their strength, restricted, however, to the cases when polarization has an out-of-plane component.

# Acknowledgements

This work was supported by the Deutsche Forschungsgemeinschaft (German Research Society, DFG) via the grant No. 405631895 (GE-1171/8-1). FEM computations for this research were conducted on the Lichtenberg high performance computer of the TU Darmstadt.

# Appendix A. Material parameters

A thin film of PZT 60/40 was considered for this investigation. Parameters were predominantly taken from works by Haun, Cross *et al.* [106], Schrade, Xu *et al.* [107], Pertsev *et al.* [108] and from the DFT calculations by Völker *et al.* [109]. Table AI shows material parameters considered in the work. Note that the tensor elements are marked using the Voigt notation.

**Table AI**. Material Parameters of PZT 60/40

| Description | Designation | Value | Units |
|---|---|---|---|
| Temperature | $T$ | 300 | °K |
| Phase-transition temperature | $T_0$ | 637.3 | °K |
| Film thickness | $h$ | 24 | nm |
| Ambience layer thickness | $h_v$ | 12 | nm |
| Background dielectric permittivity | $\varepsilon_b$ | 7 | dimensionless |
| Ambience dielectric permittivity | $\varepsilon_v$ | 1 | dimensionless |
| Surface screening length | $\lambda$ | >>1 | nm |
| Temperature-independent factor in $\alpha_1$ | $\alpha_{1T}$ | $2.331 \times 10^5$ | $m/(F \cdot K)$ |
| Landau coefficient | $\alpha_1$ | $-7.904 \times 10^7$ | $m/F$ |
| Landau coefficient | $\beta_{11}$ | $13.69 \times 10^7$ | $\dfrac{m^5}{C^2 F}$ |
| Landau coefficient | $\beta_{12}$ | $23.91 \times 10^7$ | $\dfrac{m^5}{C^2 F}$ |
| Landau coefficient | $\gamma_{111}$ | $2.713 \times 10^8$ | $\dfrac{m^9}{C^4 F}$ |
| Landau coefficient | $\gamma_{112}$ | $12.13 \times 10^8$ | $\dfrac{m^9}{C^4 F}$ |
| Landau coefficient | $\gamma_{123}$ | $-56.90 \times 10^8$ | $\dfrac{m^9}{C^4 F}$ |
| Gradient term coefficient | $g_{11}$ | $15 \times 10^{-11}$ | $m^3/F$ |
| Gradient term coefficient | $g_{12}$ | $3.5 \times 10^{-11}$ | $m^3/F$ |
| Gradient term coefficient | $g_{44}$ | $6.3 \times 10^{-11}$ | $m^3/F$ |
| Electrostriction coefficient | $Q_{11}$ | $7.260 \times 10^{-2}$ | $\dfrac{m^4}{C^2}$ |
| Electrostriction coefficient | $Q_{12}$ | $-2.708 \times 10^{-2}$ | $\dfrac{m^4}{C^2}$ |
| Electrostriction coefficient | $Q_{44}$ | $6.293 \times 10^{-2}$ | $\dfrac{m^4}{C^2}$ |
| Elastic stiffness coefficient | $c_{11}$ | $\dfrac{s_{11}+s_{12}}{(s_{11}-s_{12})(s_{11}+2s_{12})}$ | $N/m^2$ |
| Elastic stiffness coefficient | $c_{12}$ | $\dfrac{-s_{12}}{(s_{11}-s_{12})(s_{11}+2s_{12})}$ | $N/m^2$ |
| Elastic stiffness coefficient | $c_{44}$ | $\dfrac{1}{s_{44}}$ | $N/m^2$ |
| Elastic compliance coefficient | $s_{11}$ | $8.8 \times 10^{-12}$ | $m^2/N$ |
| Elastic compliance coefficient | $s_{12}$ | $-2.9 \times 10^{-12}$ | $m^2/N$ |
| Elastic compliance coefficient | $s_{44}$ | $24.6 \times 10^{-12}$ | $m^2/N$ |
| Flexoelectric coefficient | $F_{11}$ | $3 \times 10^{-11}$ | $m^3/C$ |
| Flexoelectric coefficient | $F_{12}$ | $1 \times 10^{-11}$ | $m^3/C$ |
| Flexoelectric coefficient | $F_{44}$ | $0.5 \times 10^{-11}$ | $m^3/C$ |
| Khalatnikov coefficient | $\Gamma$ | $10^5$ | $\dfrac{s \cdot m}{F}$ |

Given here are tensor components in the Voigt notation. To correctly substitute them into the energy expression (2), the following correspondence must be borne in mind:

$$s_{1111} = s_{11}, \quad s_{1122} = s_{12}, \quad 4s_{1212} = s_{44}, \tag{A.1a}$$

$$Q_{1111} = Q_{11}, \quad Q_{1122} = Q_{12}, \quad 4Q_{1212} = Q_{44}, \tag{A.1b}$$

$$F_{1111} = F_{11}, \quad F_{1122} = F_{12}, \quad 4F_{1212} = F_{44}, \tag{A.1c}$$

$$\alpha_{11} = \alpha_1, \tag{A.1d}$$

$$\beta_{1111} = \beta_{11}, \quad 6\beta_{1122} = \beta_{12}, \tag{A.1e}$$

$$\gamma_{111111} = \gamma_{111}, \quad 15\gamma_{111122} = \gamma_{112}, \quad 90\gamma_{112233} = \gamma_{123}. \tag{A.1f}$$

# Appendix B. The LGD analysis of single-domain states of thick and thin films

## a. The free energy for a homogeneous polarization

Within the continuous medium Landau-Ginzburg-Devonshire (LGD) approach [110], the value and orientation of the spontaneous polarization $P_i$ in thin ferroelectric films can be controlled by size effect, temperature $T$ and misfit strain $u_m$ originating from the film-substrate lattice constants mismatch [111]. The density of Gibbs energy, which minimization allows one to calculate the phase diagram for the homogeneous polarization, has the form (see **Appendix C** for derivation):

$$g_L = a_1 P_1^2 + a_2 P_2^2 + a_6 P_1 P_2 + a_3 P_3^2 + a_{11}(P_1^4 + P_2^4) + a_{33} P_3^4 + a_{12} P_1^2 P_2^2 + a_{13}(P_1^2 + P_2^2) P_3^2 + a_{111}(P_1^6 + P_2^6 + P_3^6) + a_{112}[P_1^2(P_2^4 + P_3^4) + P_2^2(P_1^4 + P_3^4) + P_3^2(P_2^4 + P_1^4)] + a_{123} P_1^2 P_2^2 P_3^2 \quad \text{(B1)}$$

The coefficients:

$$a_1 = \alpha_{1T}(T - T_C^f) - \frac{1}{2}\frac{\left(u_1^{(m)}+u_2^{(m)}\right)(Q_{11}+Q_{12})}{s_{11}+s_{12}} - \frac{1}{2}\frac{\left(u_1^{(m)}-u_2^{(m)}\right)(Q_{11}-Q_{12})}{s_{11}-s_{12}}, \quad \text{(B2a)}$$

$$a_2 = \alpha_{1T}(T - T_C^f) - \frac{1}{2}\frac{\left(u_1^{(m)}+u_2^{(m)}\right)(Q_{11}+Q_{12})}{s_{11}+s_{12}} + \frac{1}{2}\frac{\left(u_1^{(m)}-u_2^{(m)}\right)(Q_{11}-Q_{12})}{s_{11}-s_{12}}, \quad \text{(B2b)}$$

$$a_6 = -\frac{u_6^{(m)} Q_{44}}{s_{44}}, \quad a_3 = \alpha_{1T}(T - T_C^f) - \frac{Q_{12}\left(u_1^{(m)}+u_2^{(m)}\right)}{s_{11}+s_{12}} + \frac{d_{eff}}{2\varepsilon_0 \varepsilon_b (h+d_{eff})}, \quad \text{(B2c)}$$

$$a_{11} = \beta_{11} + \frac{s_{11}(Q_{11}^2+Q_{12}^2)-2Q_{11}Q_{12}s_{12}}{2(s_{11}^2-s_{12}^2)}, \quad a_{33} = \beta_{11} + \frac{Q_{12}^2}{s_{11}+s_{12}}, \quad \text{(B2d)}$$

$$a_{12} = \beta_{12} - \frac{s_{12}(Q_{11}^2+Q_{12}^2)-2Q_{11}Q_{12}s_{11}}{s_{11}^2-s_{12}^2} + \frac{(Q_{44})^2}{2s_{44}}, \quad a_{13} = \beta_{12} + \frac{Q_{12}(Q_{11}+Q_{12})}{s_{11}+s_{12}}. \quad \text{(B2e)}$$

$$a_{111} = \gamma_{111}, a_{112} = \gamma_{112}, a_{123} = \gamma_{123} \quad \text{(B2f)}$$

Here $T_C^f$ is the Curie temperature of bulk ferroelectric, $Q_{ij}$ are the components of electrostriction tensor, $s_{ij}$ are elastic compliances, $u_1^{(m)}$, $u_2^{(m)}$ and $u_6^{(m)}$ are components of anisotropic mismatch strain. Two diagonal components are determined by the difference of lattice constants in corresponding direction, while $u_6^{(m)}$ is the difference between the corresponding angles of the unit cells of the film and the substrate in the Voight notations. They determine the strain through the conditions at the film-substrate interface for the strain tensor $u_{ij}$, namely, $u_1 = u_1^{(m)}$ and $u_2 = u_2^{(m)}$, while $u_6 = 2u_6^{(m)}$. The Voight notations are used hereinafter (xx→1, yy→2, zz→3, zy→4, xz→5, xy→6).

The positive coefficient $\alpha_T$ is proportional to the inverse Curie-Weiss constant. When deriving Eq. (B2c) we used the expression for the depolarization field inside the ferroelectric film, $E_3 = -\frac{P_3}{\varepsilon_0 \varepsilon_b}\frac{d_{eff}}{h+d_{eff}}$ with $d_{eff} = h_v \frac{\varepsilon_b}{\varepsilon_v}$.

## b. Phase diagrams for homogeneous polarization

Phase diagrams of thick and thin PbZr$_{0.6}$Ti$_{0.4}$O$_3$ (**PZT**) films with homogeneous polarization in coordinates temperature – mismatch strain, which contain paraelectric (PE) and ferroelectric (FE) phases with out-of-plane (tetragonal FE$_C$ phase with $P_3 \neq 0, P_1 = P_2 = 0$), in-plane (orthorhombic FE$aa$ phase with $P_3 = 0, P_1 = \pm P_2 \neq 0$) and both (rhombohedral FE$r$ phase with $P_3 \neq 0, P_1 = \pm P_2 \neq 0$) orientations of polarization vector are shown in **Figs. B1a-B4a.** The dependence of the in-plane and out-of-plane polarization components on the misfit strain and temperature are shown in **Figs. B1-B4(b-d)**.

The phase designation and classification are taken from papers [15, 16, 69, 74, [112]] by Pertsev et al. **Figs. B1-2** are plotted for biaxial misfit strain, $u_1^{(m)} \equiv u_2^{(m)} = u_m$, and so the symmetry of the in-plane polarization direction, $P_1 = \pm P_2$, holds for the strain too. The symmetry $P_1 = \pm P_2$ follows from Eqs. (B2a)-(B2b), which lead to the same expressions for $a_{1,2}$ in the case, namely $a_1 = a_2 = \alpha_{1T}(T - T_C^f) - \frac{u_m(Q_{11}+Q_{12})}{s_{11}+s_{12}}$.

**Figs. B3-4** are plotted for uniaxial mismatch strain, $u_1^{(m)} \equiv u_{mX}$, while $u_2^{(m)} = 0$ and $u_6^{(m)} \equiv 0$, so the evident asymmetry $P_1 \neq \pm P_2$, exists for the case. The asymmetry can be explained as follows.

In existing perovskites with $Q_{11} > 0, Q_{12} < 0$, stretching along the $x_1$ axis corresponding to $u_1^{(m)} > 0$ leads to the appearance and enhancement of the spontaneous polarization component along the same direction, i.e., to an increase in $P_x \equiv P_1$, and its existence at higher temperatures than $P_2$. On the contrary, a compression along the $x$ axis, corresponding to $u_1^{(m)} < 0$, leads to suppression and weakening of the spontaneous polarization component along the same direction, i.e., to a decrease in $P_x \equiv P_1$, and its disappearance at lower temperatures compared to $P_2$. This effect of "asymmetry" between $P_1$ and $P_2$ is clearly seen when comparing **Fig. B3b** with **B3c** plotted for thick films and becomes even more significant for thin films (compare **Fig. B4b** with **B4c**). The asymmetry $P_1 \neq \pm P_2$ follows from Eqs. (B2a)-(B2b), which lead to different expressions for $a_{1,2}$ in the case, namely $a_1 = \alpha_{1T}(T - T_C^f) - \frac{1}{2}\frac{u_{mX}(Q_{11}+Q_{12})}{s_{11}+s_{12}} - \frac{1}{2}\frac{u_{mX}(Q_{11}-Q_{12})}{s_{11}-s_{12}}$, $a_2 = \alpha_{1T}(T - T_C^f) - \frac{1}{2}\frac{u_{mX}(Q_{11}+Q_{12})}{s_{11}+s_{12}} + \frac{1}{2}\frac{u_{mX}(Q_{11}-Q_{12})}{s_{11}-s_{12}}$, and so $a_1 - a_2 = -\frac{u_{mX}(Q_{11}-Q_{12})}{s_{11}-s_{12}}$.

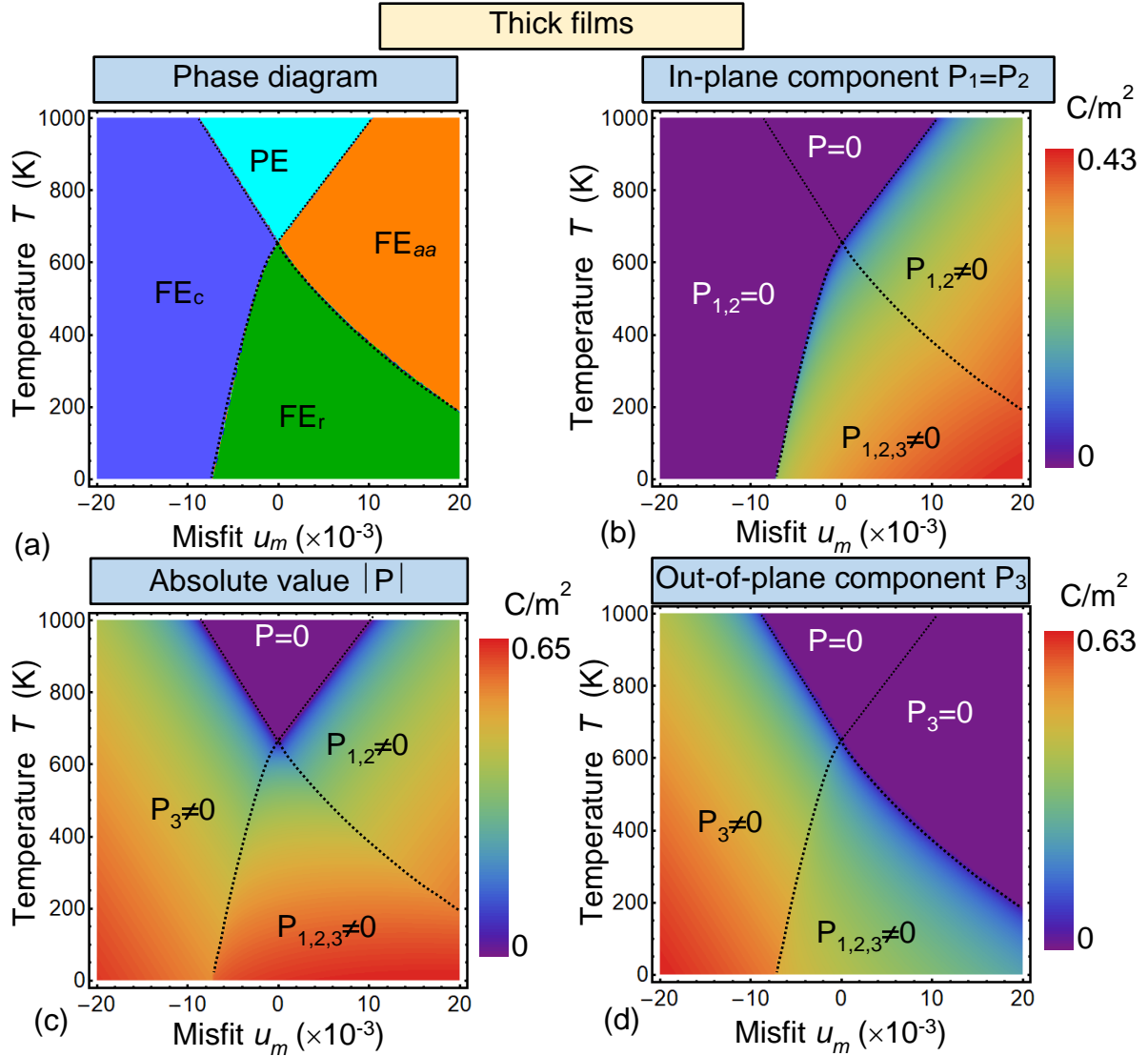

**FIGURE B1.** (a) Ferroelectric PZT 60/40 phase diagram in coordinates temperature – misfit strain for **thick films.** *Isotropic biaxial* misfit strain between the film and substrate is assumed, $u_1^{(m)} \equiv u_2^{(m)} = u_m$, while $u_6^{(m)} \equiv 0$. Paraelectric phase is denoted as "PE", while "FE$_c$", "FE$_a$" and "FE$_r$" show the regions for ferroelectric phases with out-of-plane, in-plane and mixed orientations of polarization, respectively. (b) In-plane and (d) out-of-plane components of polarization along with (c) polarization absolute value dependence on the misfit strain and temperature. Color bars show the range of corresponding values for the contour maps. Film thickness is $h > 10^4$ nm and effective spatial gap width is $d_{eff}$=0.12 nm. Bulk parameters are listed in **Table BI**.

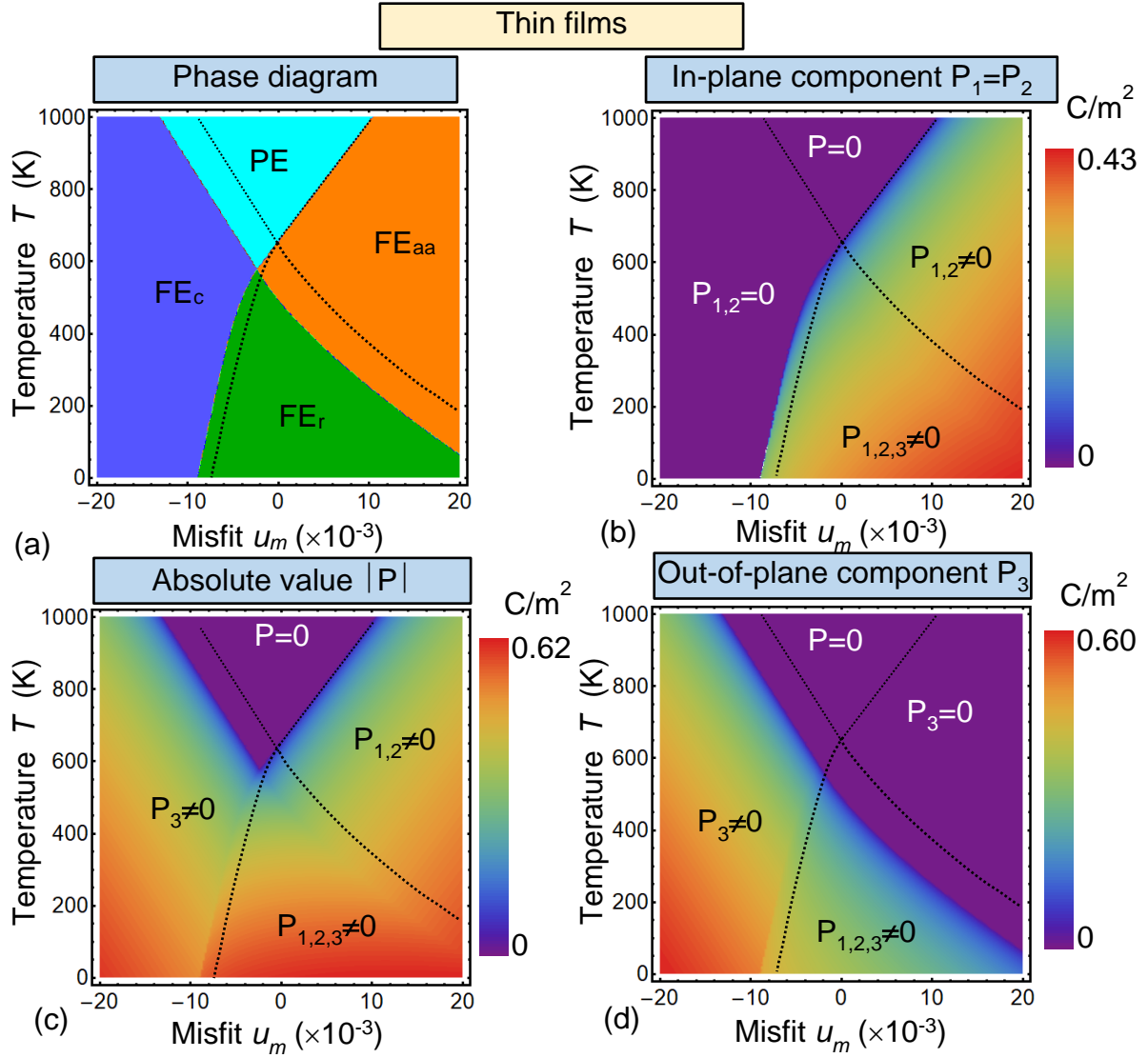

**FIGURE B2.** (a) Ferroelectric PZT 60/40 phase diagram in coordinates temperature – misfit strain for **thin films**. *Isotropic biaxial* misfit strain between the film and substrate is assumed, $u_1^{(m)} \equiv u_2^{(m)} = u_m$, while $u_6^{(m)} \equiv 0$. Paraelectric phase is denoted as "PE", while "FE$_c$", "FE$_a$" and "FE$_r$" show the regions for ferroelectric phases with out-of-plane, in-plane and mixed orientations of polarization, respectively. (b) In-plane and (d) out-of-plane components of polarization along with (c) polarization absolute value dependence on the misfit and temperature. Color bars show the range of corresponding values for the contour maps. Film thickness is $h=24$ nm and effective spatial gap width is $d_{eff}=0.12$ nm. Bulk parameters are listed in **Table BI**.

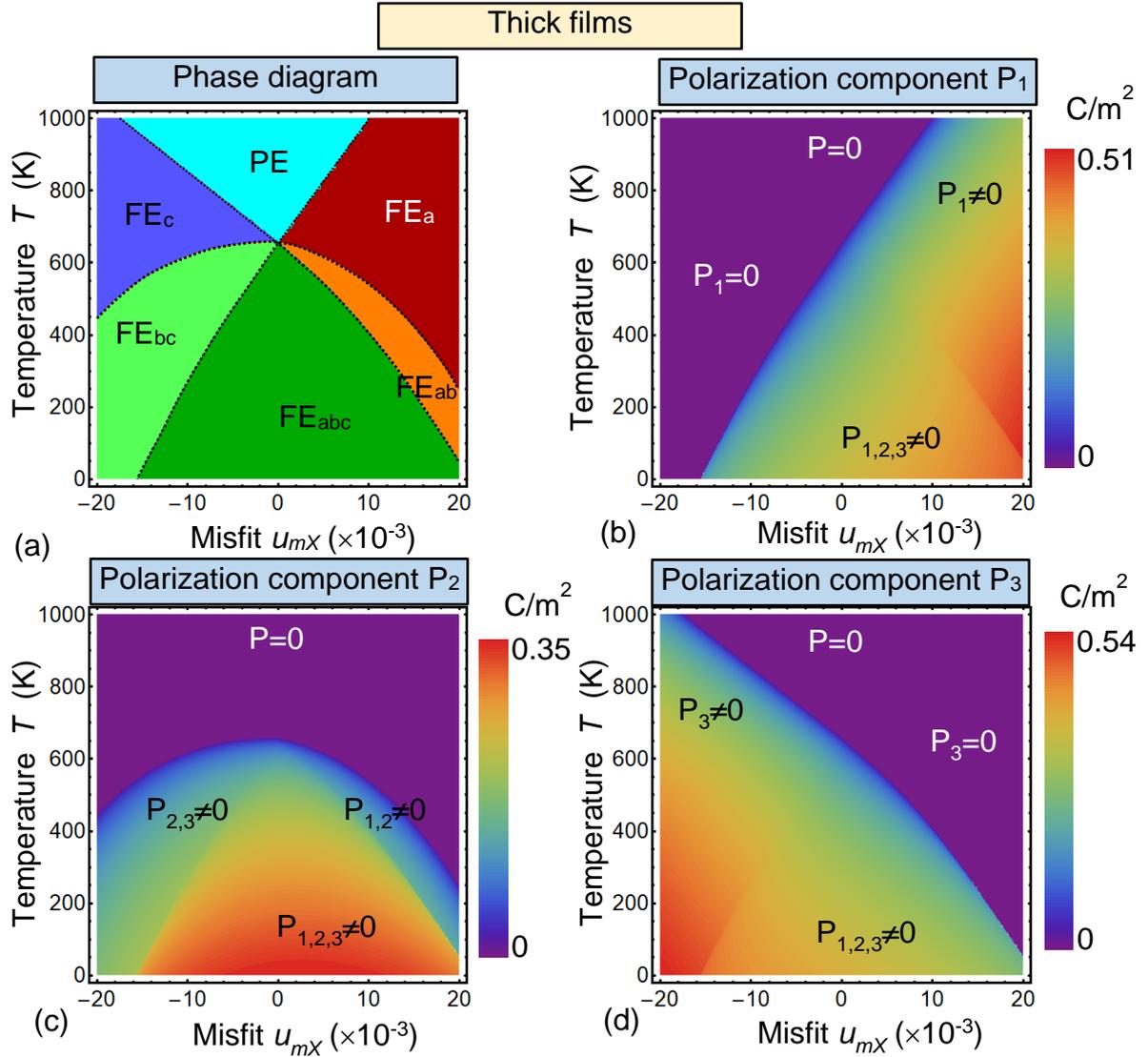

**FIGURE B3.** (a) Ferroelectric PZT 60/40 phase diagram in coordinates temperature – misfit strain for **thick films.** *Anisotropic uniaxial* misfit strain between the film and substrate is assumed, $u_1^{(m)} \equiv u_{mX}$, while $u_2^{(m)} = 0$ and $u_6^{(m)} \equiv 0$. Polarization components $P_1$ (b), $P_2$ (c) and $P_3$ (d) dependence on the misfit strain and temperature. Color bars show the range of corresponding values for the contour maps. Paraelectric phase is denoted as "PE", while "FE$_a$", "FE$_{ab}$", "FE$_{abc}$", "FE$_{bc}$", and "FE$_c$" show the regions for ferroelectric phases with different orientations of polarization, namely, "a" means $P_1$, "b" means $P_2$ and "c" means $P_3$. Film thickness is $h > 10^4$ nm and effective spatial gap width is $d_{eff} = 0.12$ nm. Bulk parameters are listed in **Table BI.**

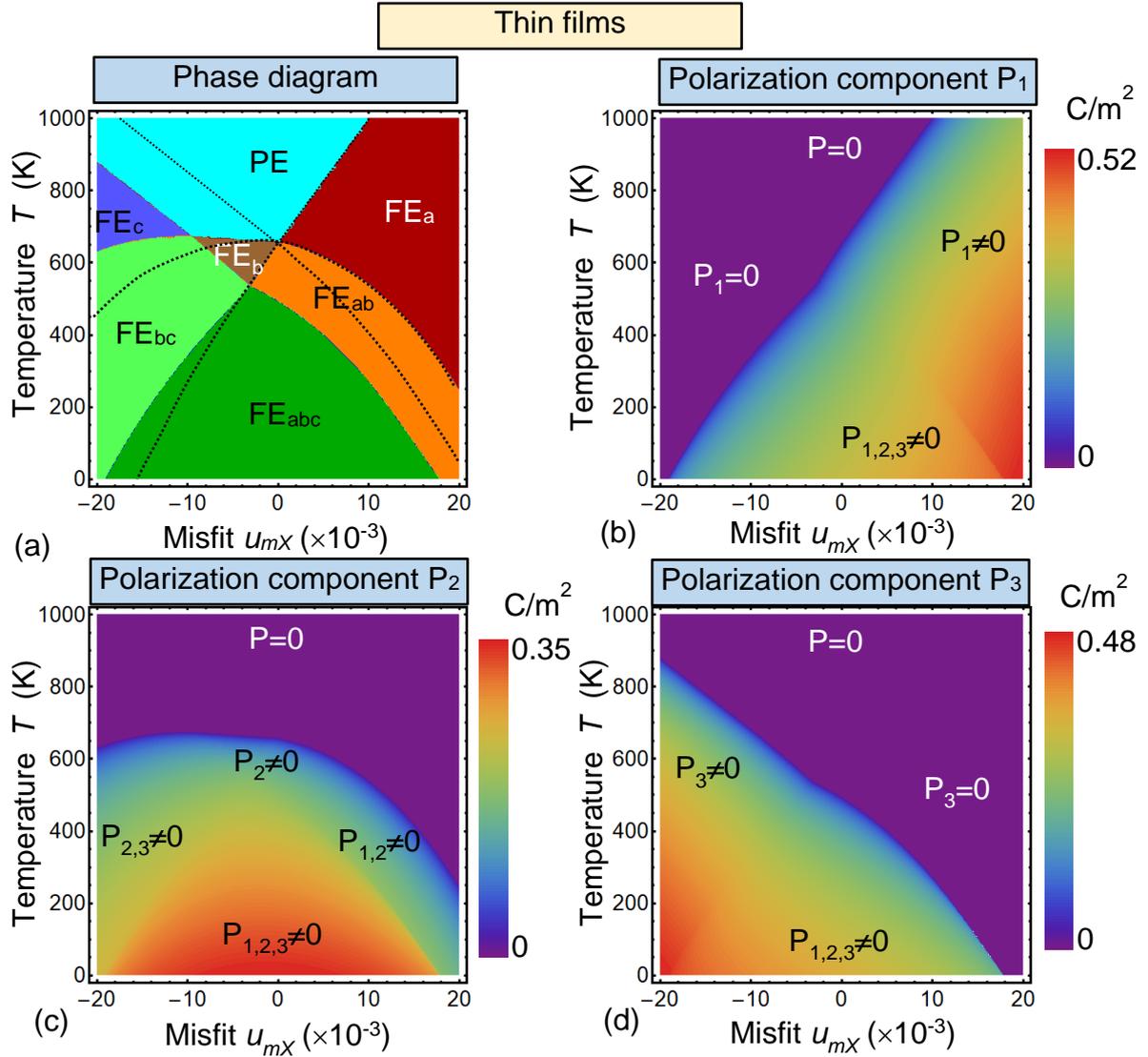

**FIGURE B4**. **(a)** Ferroelectric PZT 60/40 phase diagram in coordinates temperature – misfit strain for **thin films.** *Anisotropic uniaxial* misfit strain between the film and substrate is assumed, $u_1^{(m)} \equiv u_{mX}$, while $u_2^{(m)} = 0$ and $u_6^{(m)} \equiv 0$. Polarization components $P_1$ **(b)**, $P_2$ **(c)** and $P_3$ **(d)** dependence on the misfit strain and temperature. Color bars show the range of corresponding values for the contour maps. Paraelectric phase is denoted as "PE", while "FE$_a$", "FE$_b$", "FE$_{ab}$", "FE$_{abc}$", "FE$_{bc}$", and "FE$_c$" show the regions for ferroelectric phases with different orientations of polarization, namely, "a" means $P_1$, "b" means $P_2$ and "c" means $P_3$. Film thickness is $h=24$ nm and effective spatial gap width is $d_{eff}=0.12$ nm. Bulk parameters are listed in **Table BI**.

# Appendix C. The LGD approach for an anisotropic misfit strain in [001] oriented films (pseudo 4mm symmetry)

Considering uniform polarization states and thus neglecting the gradient contributions to the energy the elastic Gibbs potential (2a) in the absence of the electric field can be reduced to the form

$$\Delta G_{FE} = a_1 (P_1^2 + P_2^2 + P_3^2) + a_{11}(P_1^4 + P_2^4 + P_3^4) + a_{12}(P_1^2 P_2^2 + P_2^2 P_3^2 + P_1^2 P_3^2) +$$
$$a_{111}(P_1^6 + P_2^6 + P_3^6) + a_{112}(P_1^4(P_2^2 + P_3^2) + (P_1^2 + P_3^2)P_2^4 + (P_1^2 + P_2^2)P_3^4) + a_{123} P_1^2 P_2^2 P_3^2 -$$
$$Q_{11}(\sigma_1 P_1^2 + \sigma_2 P_2^2 + \sigma_3 P_3^2) - Q_{12}(\sigma_1 P_2^2 + \sigma_2 P_1^2 + \sigma_1 P_3^2 + \sigma_3 P_1^2 + \sigma_2 P_3^2 + \sigma_3 P_2^2) -$$
$$Q_{44}(\sigma_6 P_1 P_2 + \sigma_5 P_1 P_3 + \sigma_4 P_2 P_3) - \frac{s_{11}}{2}(\sigma_1^2 + \sigma_2^2 + \sigma_3^2) - s_{12}(\sigma_1 \sigma_2 + \sigma_1 \sigma_3 + \sigma_2 \sigma_3) -$$
$$\frac{s_{44}}{2}(\sigma_4^2 + \sigma_5^2 + \sigma_6^2). \quad \text{(C.1)}$$

Here Voight matrix notations are used in Eq. (C.1)

$$s_{1111} = s_{11}, \quad s_{1122} = s_{12}, \quad 4s_{1212} = s_{44}, \quad \text{(C.2a)}$$

$$Q_{1111} = Q_{11}, \quad Q_{1122} = Q_{12}, \quad 4Q_{1212} = Q_{44}, \quad \text{(C.2b)}$$

$$a_{1111} = a_{11}, \quad 6a_{1122} = a_{12}, \quad a_{111111} = a_{111}, \quad 15a_{111122} = a_{112}, \quad 90a_{112233} = a_{123}, \quad \text{(C.2c)}$$

$$\sigma_{11} = \sigma_1, \quad \sigma_{22} = \sigma_2, \quad \sigma_{33} = \sigma_3, \quad \sigma_{23} = \sigma_4, \quad \sigma_{13} = \sigma_5, \quad \sigma_{12} = \sigma_6. \quad \text{(C.2d)}$$

$$u_{11} = u_1, \quad u_{22} = u_2, \quad u_{33} = u_3, \quad u_{23} = u_4, \quad u_{13} = u_5, \quad u_{12} = u_6. \quad \text{(C.2e)}$$

Modified Hooke's law could be obtained from the relation $u_i = -\partial(\Delta G_{FE})/\partial \sigma_i$:

$$u_1 = s_{11}\sigma_1 + s_{12}\sigma_2 + s_{12}\sigma_3 + Q_{11} P_1^2 + Q_{12} P_2^2 + Q_{12} P_3^2 \quad \text{(C.3a)}$$

$$u_2 = s_{12}\sigma_1 + s_{11}\sigma_2 + s_{12}\sigma_3 + Q_{12} P_1^2 + Q_{11} P_2^2 + Q_{12} P_3^2 \quad \text{(C.3b)}$$

$$u_3 = s_{12}\sigma_1 + s_{12}\sigma_2 + s_{11}\sigma_3 + Q_{12} P_1^2 + Q_{12} P_2^2 + Q_{11} P_3^2 \quad \text{(C.3c)}$$

$$u_4 = s_{44}\sigma_4 + Q_{44} P_2 P_3 \quad \text{(C.3d)}$$

$$u_5 = s_{44}\sigma_5 + Q_{44} P_1 P_3 \quad \text{(C.3e)}$$

$$u_6 = s_{44}\sigma_6 + Q_{44} P_1 P_2 \quad \text{(C.3f)}$$

The solution for the misfit of a thin film with its substrate is well known. For the film with the normal along $x_3$ direction one has the following relations for stress and strain components (see e.g. [15]):

$$\sigma_4 = \sigma_5 = \sigma_3 = 0 \quad \text{(C.4a)}$$

$$u_1 = u_1^{(m)}, \quad u_2 = u_2^{(m)}, \quad u_6 = u_6^{(m)} \quad \text{(C.4b)}$$

Here the definitions of misfit strain components $u_1^{(m)}$, $u_2^{(m)}$ and $u_6^{(m)}$ are given after Eqs.(B.2e). Taking (C.3) and (C.4) into account

$$u_1^{(m)} = s_{11}\sigma_1 + s_{12}\sigma_2 + Q_{11}P_1^2 + Q_{12}P_2^2 + Q_{12}P_3^2 \tag{C.5a}$$

$$u_2^{(m)} = s_{12}\sigma_1 + s_{11}\sigma_2 + Q_{12}P_1^2 + Q_{11}P_2^2 + Q_{12}P_3^2 \tag{C.5b}$$

$$u_3 = s_{12}\sigma_1 + s_{12}\sigma_2 + Q_{12}P_1^2 + Q_{12}P_2^2 + Q_{11}P_3^2 \tag{C.5c}$$

$$u_4 = Q_{44}P_2P_3, \quad u_5 = Q_{44}P_1P_3, \quad u_6^{(m)} = s_{44}\sigma_6 + Q_{44}P_1P_2. \tag{C.5d}$$

The solution of the system (C.5) is

$$\sigma_1 = \frac{s_{11}\delta u_1 - s_{12}\delta u_2}{s_{11}^2 - s_{12}^2} \equiv \frac{1}{2}\left(\frac{\delta u_1 + \delta u_2}{s_{11} + s_{12}} + \frac{\delta u_1 - \delta u_2}{s_{11} - s_{12}}\right), \tag{C.6a}$$

$$\sigma_2 = \frac{s_{11}\delta u_2 - s_{12}\delta u_1}{s_{11}^2 - s_{12}^2} \equiv \frac{1}{2}\left(\frac{\delta u_1 + \delta u_2}{s_{11} + s_{12}} - \frac{\delta u_1 - \delta u_2}{s_{11} - s_{12}}\right), \tag{C.6b}$$

$$u_3 = s_{12}\frac{\delta u_1 + \delta u_2}{s_{11} + s_{12}} + Q_{12}P_1^2 + Q_{12}P_2^2 + Q_{11}P_3^2, \tag{C.6c}$$

$$\sigma_6 = \frac{u_6^{(m)} - Q_{44}P_1P_2}{s_{44}}, \quad u_4 = Q_{44}P_2P_3, \quad u_5 = Q_{44}P_1P_3. \tag{C.6d}$$

Here we introduce values $\delta u_1$ and $\delta u_2$ as the differences between misfit and spontaneous strain components.

$$\delta u_1 = u_1^{(m)} - Q_{11}P_1^2 - Q_{12}P_2^2 - Q_{12}P_3^2, \tag{C.6e}$$

$$\delta u_2 = u_2^{(m)} - Q_{12}P_1^2 - Q_{11}P_2^2 - Q_{12}P_3^2, \tag{C.6f}$$

Finally, the Legendre transformation of the expression (C.1) for Gibbs potential gives us the Helmholtz free energy:

$$\widetilde{\Delta F}_{FE} = \Delta G_{FE} + \sigma_1 u_1 + \sigma_2 u_2 + \sigma_6 u_6 = a_1^{(P)}(P_1^2 + P_2^2 + P_3^2) + a_{11}^{(P)}(P_1^4 + P_3^4) + a_{12}^{(P)}(P_1^2 P_2^2 + P_1^2 P_3^2 + P_2^2 P_3^2) + a_{111}^{(P)}(P_1^6 + P_3^6) + a_{112}^{(P)}(P_1^4 P_3^2 + P_1^2 P_3^4) - Q_{11}(\sigma_1 P_1^2 + \sigma_{22} P_2^2) - Q_{12}\big(\sigma_1(P_2^2 + P_3^2) + \sigma_2(P_1^2 + P_3^2)\big) - Q_{44}\sigma_4 P_1 P_2 - \frac{s_{11}}{2}(\sigma_1^2 + \sigma_2^2) - s_{12}\sigma_1 \sigma_2 - \frac{s_{44}}{2}\sigma_6^2 + \sigma_1 u_1^{(m)} + \sigma_2 u_2^{(m)} + \sigma_6 u_6^{(m)}, \tag{C.7a}$$

where an "elastic" part of (C.7a) can be transformed as follows,

$$-Q_{11}(\sigma_1 P_1^2 + \sigma_2 P_2^2) - Q_{12}\big(\sigma_1(P_2^2 + P_3^2) + \sigma_2(P_1^2 + P_3^2)\big) - Q_{44}\sigma_6 P_1 P_2 - \frac{s_{11}}{2}(\sigma_1^2 + \sigma_2^2) - s_{12}\sigma_1\sigma_2 - \frac{s_{44}}{2}\sigma_6^2 + \sigma_1 u_1^{(m)} + \sigma_2 u_2^{(m)} = \sigma_1 \delta u_1 + \sigma_2 \delta u_2 - \frac{s_{11}}{2}(\sigma_1^2 + \sigma_2^2) - s_{12}\sigma_1\sigma_2 - \frac{s_{44}}{2}\sigma_6^2 + \sigma_{12}\left(u_6^{(m)} - Q_{44}P_1 P_2\right) = \frac{(\sigma_1 + \sigma_2)(\delta u_1 + \delta u_2)}{2} + \frac{(\sigma_1 - \sigma_2)(\delta u_1 - \delta u_2)}{2} - \frac{(s_{11} + s_{12})}{4}(\sigma_1 + \sigma_2)^2 - \frac{(s_{11} - s_{12})}{4}(\sigma_1 - \sigma_2)^2 + \frac{\left(u_6^{(m)} - Q_{44}P_1 P_2\right)^2}{2s_{44}} = \frac{(\delta u_1 + \delta u_2)^2}{4(s_{11}+s_{12})} + \frac{(\delta u_1 - \delta u_2)^2}{4(s_{11}-s_{12})} + \frac{\left(u_6^{(m)} - Q_{44}P_1 P_2\right)^2}{2s_{44}} = \frac{\left(u_1^{(m)} + u_2^{(m)} - (Q_{11}+Q_{12})(P_1^2+P_2^2) - 2Q_{12}P_3^2\right)^2}{4(s_{11}+s_{12})} + \frac{\left(u_1^{(m)} - u_2^{(m)} - (Q_{11}-Q_{12})(P_1^2-P_2^2)\right)^2}{4(s_{11}-s_{12})} + \frac{\left(u_6^{(m)} - Q_{44}P_1 P_2\right)^2}{2s_{44}} \tag{C.7b}$$

After all, the free energy due to misfit strain acquires the form

$$\widetilde{\Delta F}_{FE} = \frac{\left(u_1^{(m)}+u_2^{(m)}\right)^2}{4(s_{11}+s_{12})} + \frac{\left(u_1^{(m)}-u_2^{(m)}\right)^2}{4(s_{11}-s_{12})} + \frac{\left(u_6^{(m)}\right)^2}{2s_{44}} + \left(a_1^{(P)} - \frac{1}{2}\frac{\left(u_1^{(m)}+u_2^{(m)}\right)(Q_{11}+Q_{12})}{s_{11}+s_{12}} - \right.$$
$$\left. \frac{1}{2}\frac{\left(u_1^{(m)}-u_2^{(m)}\right)(Q_{11}-Q_{12})}{s_{11}-s_{12}}\right)P_1^2 + \left(a_1^{(P)} - \frac{1}{2}\frac{\left(u_1^{(m)}+u_2^{(m)}\right)(Q_{11}+Q_{12})}{s_{11}+s_{12}} + \frac{1}{2}\frac{\left(u_1^{(m)}-u_2^{(m)}\right)(Q_{11}-Q_{12})}{s_{11}-s_{12}}\right)P_2^2 -$$
$$\frac{u_6^{(m)}Q_{44}}{s_{44}}P_1P_2 + \left(a_1^{(P)} - \frac{\left(u_1^{(m)}+u_2^{(m)}\right)Q_{12}}{s_{11}+s_{12}}\right)P_3^2 + \left(a_{11}^{(P)} + \frac{1}{4}\frac{(Q_{11}+Q_{12})^2}{s_{11}+s_{12}} + \frac{1}{4}\frac{(Q_{11}-Q_{12})^2}{s_{11}-s_{12}}\right)(P_1^4 + P_2^4) +$$
$$\left(a_{11}^{(P)} + \frac{(Q_{12})^2}{s_{11}+s_{12}}\right)P_3^4 + \left(a_{12}^{(P)} + \frac{1}{2}\frac{(Q_{11}+Q_{12})^2}{s_{11}+s_{12}} - \frac{1}{2}\frac{(Q_{11}-Q_{12})^2}{s_{11}-s_{12}} + \frac{(Q_{44})^2}{2s_{44}}\right)P_1^2P_2^2 + \left(a_{12}^{(P)} + \right.$$
$$\left.\frac{(Q_{11}+Q_{12})Q_{12}}{s_{11}+s_{12}}\right)(P_1^2+P_2^2)P_3^2 + a_{111}^{(P)}(P_1^6+P_2^6+P_3^6) + a_{112}^{(P)}(P_1^4(P_2^2+P_3^2) + (P_1^2+P_3^2)P_2^4 +$$
$$(P_1^2 + P_2^2)P_3^4) + a_{123}^{(P)}P_1^2 P_2^2 P_3^2 \qquad (C.8)$$

Analysis of the coefficients before $P_i^2$, $P_1P_2$, $P_i^2 P_j^2$, and $P_i^4$, leads to Eqs. (B.2).

## Appendix D. Flexoelectric Impact on Domain Structure

Displayed in Fig. 4 polarization vector maps allow observation of domain structures within the film, and their changes due to the application of misfit strains. The other factor that also affects the form of domains is the presence of the flexoelectric effect. Though not of the critical nature, the flexoelectric impact at compressive misfit strains is still visible quantitatively. To help deduce important parameters quantifying that influence, analytical calculations were conducted.

The impact of the flexoelectricity and gradient effects within the Gibbs free energy density is presented by following terms:

$$g_{LGD} = g_L + g_{grad} + g_{flexo}, \quad g_{grad} = \frac{g_{ijkl}}{2}\frac{\partial P_i}{\partial x_j}\frac{\partial P_k}{\partial x_l}, \quad g_{flexo} = \frac{f_{ijkl}}{2}u_{ij}\frac{\partial P_k}{\partial x_l}, \qquad (11)$$

where $g_L$ is given by the first three terms in Eq. (2a), the elastic strains are given by Eq. (8) and full matrix notations are used. The flexoelectric contribution in terms of strain $u_{ij}$ and the corresponding flexoelectric tensor $f_{ijkl}$ can be derived from the last term in Eq. (2a) by a Legendre transformation. The substitution of the electrostrictive part of the strains (a toy model for which the term $s_{ijkl}\sigma_{kl}$ can be regarded small) leads to:

$$\delta g_{grad+flexo} = \frac{g_{ijkl}}{2}\frac{\partial P_i}{\partial x_j}\frac{\partial P_k}{\partial x_l} + \frac{f_{ijkl}}{2}\left(Q_{ijmn}P_mP_n + F_{ijmn}\frac{\partial P_m}{\partial x_n}\right)\frac{\partial P_k}{\partial x_l}$$
$$\cong \left(\frac{g_{ijkl}}{2} + \frac{f_{qskl}}{2}F_{qsij}\right)\frac{\partial P_i}{\partial x_j}\frac{\partial P_k}{\partial x_l} \equiv \frac{1}{2}\left(g_{ijkl} + s_{qsmn}f_{qskl}f_{mnij}\right)\frac{\partial P_i}{\partial x_j}\frac{\partial P_k}{\partial x_l} \equiv \frac{1}{2}g'_{ijkl}\frac{\partial P_i}{\partial x_j}\frac{\partial P_k}{\partial x_l}. \quad (12)$$

Since the term $\frac{f_{ijkl}}{2}Q_{ijmn}P_mP_n\frac{\partial P_k}{\partial x_l}$ has (almost) zero average, it can be omitted, and then the renormalized gradient coefficient $g'_{ijkl} = g_{ijkl} + s_{qsmn}f_{qskl}f_{mnij}$ can be introduced. The renormalization has different signs for the diagonal and non-diagonal components, but for the cases of interest, $g'_{11} = g_{11} + s_{qqmm}f_{qq11}f_{mm11}$ and $g'_{44} = g_{44} + s_{qqmm}f_{qq44}f_{mm44}$, it (typically) increases $g_{11}$ and decreases $g_{44}$. For a cubic symmetry of the parent phase, the trend

$g'_{11} > g_{11}$ and $g'_{44} < g_{44}$ is responsible for the increase of intrinsic width of the charged domain walls/structures/configurations and decrease of the width of uncharged domain walls. Since the uncharged walls are the most common, being significantly more preferable from the energetic viewpoint [113, 114], the flexoelectricity induces and facilitates their curvature, meandering and labyrinthine configurations in multiaxial ferroelectrics at $g'_{ijkl} < g^{cr}_{ijkl}$ (see e.g. [95, 115, 116]). At the same time, the condition $g'_{44} < g_{44}$ increases, while not very strongly, the transition temperature from the ferroelectric to paraelectric phase (see e.g. [17, 117]). The strong influence of the flexoelectricity comes from the flexo-dependent boundary conditions in strained thin films, $\left( g_{kjim} n_k \frac{\partial P_m}{\partial x_j} + a^S_{ij} P_j - F_{jkim} \sigma_{jk} n_m \right)\bigg|_{x_3=h} = 0$ (see e.g. Ref. [117] for details). The term $F_{jkim}\sigma_{jk}n_m$, proportional to the flexoelectric coupling strength, can lead to the appearance of built-in inhomogeneous flexo-electric fields with specific geometry.

Actually, due to the flexoelectric effect, there is no size-induced transition to a paraelectric phase down to (2 – 4) nm thickness of PbTiO$_3$ [001]-oriented films with 1% of compressive misfit strain [117]. The origin of this phenomenon is the re-building of the domain structure in the film (namely the cross-over from *c*-domain stripes to *a*-type closure domains) emerging with its thickness decrease below 4 nm, and conditioned by the flexoelectric coupling.